\pdfoutput=1
\documentclass[letterpaper,twocolumn,10pt]{article}
\usepackage{usenix}

\usepackage{hhline}
\usepackage{graphicx}
\usepackage{tikz}
\usetikzlibrary{shapes.multipart, positioning, calc, patterns, patterns.meta, fit, automata,arrows.meta}
\usepackage{array}
\usepackage{siunitx}
\usepackage{xparse}
\usepackage{stmaryrd}
\usepackage{amssymb}
\usepackage{amsthm}
\usepackage{amsmath}
\usepackage[fixamsmath=false]{mathtools} 

\theoremstyle{plain}
\newtheorem{theorem}{Theorem}[section]

\newtheorem{corollary}[theorem]{Corollary}
\newtheorem{remark}[theorem]{Remark}
\theoremstyle{definition}
\newtheorem{example}[theorem]{Example}

\usepackage[capitalize]{cleveref}
\crefname{remark}{remark}{remarks}
\Crefname{remark}{Remark}{Remarks}

\hypersetup{
  pdftitle={Zero-Knowledge Model Checking},
  pdfauthor={Pascal Berrang, Mirco Giacobbe, Jacob Swales, Xiao Yang}
}

\newcommand{\rchoose}{\overset{R}{\leftarrow}}

\providecommand{\bbbp}{\mathbb{P}}
\providecommand{\bbbn}{\mathbb{N}}
\providecommand{\bbbf}{\mathbb{F}}

\providecommand{\cal}{\mathcal}
\providecommand{\llangle}{\langle\!\langle}
\providecommand{\rrangle}{\rangle\!\rangle}

\newcommand{\andimplies}{%
  \mathrel{\makebox[\widthof{$\implies$}][c]{$\land$}}%
}

\def\lnext{\mathsf{X}}
\def\until{\mathsf{U}}
\def\globally{\mathsf{G}}
\def\finally{\mathsf{F}}

\def\T{\mathsf{T}}
\def\pub{\mathsf{p}}
\def\sec{\mathsf{s}}
\def\pok{\mathsf{PoK}}

\newcommand{\centeredtag}{%
  \refstepcounter{equation}%
  \tag*{\raisebox{-.5\baselineskip}{\normalfont(\theequation)}}%
}

\newcommand{\artifactlink}{\url{https://github.com/aiverification/zkmc}}

\newmuskip\savedthickmuskip
\newmuskip\savedmedmuskip
\newcommand{\tightmath}{%
  \savedthickmuskip=\thickmuskip
  \savedmedmuskip=\medmuskip
  \thickmuskip=3.5mu plus 1.5mu minus 1mu
  \medmuskip=2mu plus 1mu minus 1mu
}
\newcommand{\normalmath}{%
  \thickmuskip=\savedthickmuskip
  \medmuskip=\savedmedmuskip
}

\ExplSyntaxOn

\NewDocumentCommand{\zkroundtime}{m}
 {
  \tl_if_blank:nTF { #1 } { -- } { \zk_round_time_fp:n { #1 } }
 }
\cs_new_protected:Nn \zk_round_time_fp:n
 {
  \fp_set:Nn \l_tmpa_fp { #1 } 
  \fp_compare:nTF { \l_tmpa_fp < 1 }
   {
     \int_set:Nn \l_tmpa_int
      { \fp_to_int:n { round( 1000 * \l_tmpa_fp ) } }
     \SI{\int_use:N \l_tmpa_int}{\milli\second}
   }
   {
     \fp_compare:nTF { \l_tmpa_fp < 60 }
      {
        \int_set:Nn \l_tmpa_int
         { \fp_to_int:n { round( \l_tmpa_fp ) } }
        \SI{\int_use:N \l_tmpa_int}{\second}
      }
      {
        \int_set:Nn \l_tmpb_int
         { \fp_to_int:n { round( \l_tmpa_fp / 60 ) } }
        \int_compare:nNnTF { \l_tmpb_int } < { 120 }
         {
           \SI{\int_use:N \l_tmpb_int}{\minute}
         }
         {
           \int_set:Nn \l_tmpa_int
            { \fp_to_int:n { round( \l_tmpa_fp / 3600 ) } }
           \SI{\int_use:N \l_tmpa_int}{\hour}
         }
      }
   }
 }

\ExplSyntaxOff

\ExplSyntaxOn
\NewDocumentCommand{\logtwo}{m}
  {
    \fp_eval:n { round( ln(#1) / ln(2), 1 ) }
  }
\ExplSyntaxOff

\newcommand{\zkroundspace}[1]{$2^{\logtwo{#1}}$}

\begin{document}

\date{}

\title{\Large \bf Zero-Knowledge Model Checking}

\author{
{\rm Pascal Berrang}\\
University of Birmingham\\
Zeroth Research\\
p.p.berrang@bham.ac.uk
\and
{\rm Mirco Giacobbe}\\
University of Birmingham\\
Zeroth Research\\
m.giacobbe@bham.ac.uk
\and
{\rm Jacob Swales}\\
University of Birmingham\\
jxs1545@student.bham.ac.uk
\and
{\rm Xiao Yang}\\
University of Birmingham\\
Xi'an Jiaotong-Liverpool University\\
xiao.yang@xjtlu.edu.cn
} 

\maketitle

\begin{abstract}
  We introduce a technology to formally verify that a software system satisfies a temporal specification of functional correctness, without revealing the system itself. Our method combines a deductive approach to model checking to obtain a formal certificate of correctness for the system, with zero-knowledge proofs to convince an external verifier that the system---kept secret---complies with its specification of correctness---made public. We consider proof certificates represented as ranking functions, and introduce both an explicit-state and a symbolic scheme for model checking in zero knowledge. Our explicit-state scheme assumes systems represented as transition graphs. We use polynomial commitments to convince the verifier that the public proof certificates correspond to the secret transition relation. Our symbolic scheme assumes systems specified as linear guarded commands. We apply Farkas' lemma to obtain a witness for the validity of the ranking function
  and employ sigma protocols with folding to efficiently convince the verifier of the witness's existence.
  We built a prototype to demonstrate the practical efficacy of our two schemes on linear temporal logic verification examples. Our technology enables formal verification in domains where both the safety and the confidentiality of the system under analysis are critical.
\end{abstract}

\section{Introduction}
Model checking is the problem of determining whether a software, hardware, or cyber-physical system satisfies a formal specification of correctness. Unlike traditional testing, which is inherently non-exhaustive, model checking provides formal proof that the system behaves as intended~\cite{alur2015principles,DBLP:books/daglib/0020348,DBLP:books/daglib/0007403-2}. It is an established standard in electronic design automation, protocol verification, and control software development in domains where safety and security are highly critical. Moreover, it is increasingly viewed as a promising avenue for the safety assurance of AI systems. Formal proofs of correctness enable verifiable compliance with safety standards and foster trust between technology providers, regulators, and users.

Scalability has been the central focus of formal methods researchers for decades, yet the industrial adoption of model checking faces another pressing challenge: confidentiality.
The traditional formal verification algorithms require disclosure of the system description in order to convince an external verifier of the system's compliance with its specification.
In a setting where formal verification is a requirement potentially involving multiple organisations, jurisdictions, standardisation bodies, the obligation to disclose a system description is likely to create tension between regulators' or customers' need to ensure compliance with safety requirements and providers' desire to protect their intellectual property.

Zero-knowledge proofs provide a cryptographic mechanism by which a prover can convince a verifier that a given statement is true without revealing any information beyond the validity of the statement itself. They are widely used in areas such as blockchain systems, privacy-preserving authentication, secure electronic voting, and confidential data access control. In the context of system verification, they offer the opportunity to demonstrate compliance with formal specifications while preserving confidentiality, as correctness claims can be validated without disclosing system internals or proprietary designs.
Despite their maturity and broad adoption in multiple domains, the application of zero-knowledge proofs to formal verification remains rare~\cite{LuoAHPTW22, LuickKAHPPT0L24,DBLP:conf/ccs/LauferOB24}.

We introduce the first approach to model checking in zero knowledge: {\em Zero-Knowledge Model Checking} (ZKMC). Our method combines proof-theoretic model checking with zero-knowledge proofs to construct a proof certificate that the system satisfies the specification and then convince an external verifier that our {\em public} artefacts---the specification and the proof certificate---are valid with respect to our {\em secret} artefact---the system under analysis---without revealing any information about the system itself.

We consider the model checking of reactive systems with discrete state spaces against $\omega$-regular specifications (\cref{sec:ltl}), which subsume linear temporal logic (LTL)~\cite{DBLP:conf/focs/Pnueli77}.
Following the automata-theoretic approach~\cite{Vardi91,DBLP:conf/lics/VardiW86}, we reduce model checking to the language emptiness of the synchronous composition of the system with a B\"uchi automaton for the complementary specification; a certificate of this emptiness serves as our certificate of correctness.
On this premise, we realise ZKMC through two complementary algorithms: one explicit-state, one symbolic.

Our explicit-state algorithm (\cref{sec:explicit}) assumes systems represented as finite-state transition graphs whose states are labelled with atomic propositions, i.e., as Kripke structures. We associate the system with a proof certificate over its state space alongside an appropriate automaton. The task is to convince an external verifier of the validity of the proof certificate without revealing the system's initial states or its transition relation.  Our scheme employs KZG polynomial commitments with batch evaluation
to ensure that any hypothetical state or transition that would violate the publicly disclosed proof certificate cannot correspond to an actual initial state or transition of the secret system. Upon successful verification, which is guaranteed by the standard cryptographic properties of polynomial commitments~\cite{KateZG10}, the proof of correctness follows by contraposition.
Our proof is two group elements regardless of state-space size; prover work is quadratic in the number of states of the system.

Our symbolic algorithm (\cref{sec:symbolic}) assumes reactive systems represented over integer variables by guarded commands with linear constraints and integer coefficients, potentially yielding very large or infinite state spaces.
We require state propositions to be expressed as Boolean combinations of linear inequalities, and ranking functions as piecewise linear functions; from these, we derive proof obligations in a uniform form, each with a secret component corresponding to the coefficients of the system.
The task is to convince an external verifier that every obligation holds without revealing its secret component.
Our scheme computes a Farkas witness for each obligation and employs specialised sigma sub-protocols for the relevant operations over two-tier Pedersen commitments~\cite{CongYY24}; it then folds all proofs into a single instance, amortising the cost over the whole, possibly very large, batch of obligations.
Upon successful verification, we obtain a zero-knowledge proof of correctness for the system under analysis.
Prover and verifier cost are independent of the state-space size and polynomial in the number of guarded commands and automaton transitions.

We developed a prototype implementation and evaluated our two algorithms on a range of liveness verification tasks (\Cref{sec:eval}). Our benchmarks include a 3-way handshake protocol with exponential backoff, a model of DHCP, and a round-robin scheduling algorithm.
Our explicit-state algorithm can verify systems with up to approximately $2^{12}$ states within a \SI{2}{\hour} time limit. Under the same time constraint, our symbolic algorithm handles significantly larger tasks, scaling to approximately $2^{21}$ states. We emphasise that the explicit-state algorithm makes no assumptions about the structure of the system, whereas the symbolic algorithm requires the system to be expressed as linear guarded commands. These results therefore highlight a trade-off between expressivity and performance, with the explicit-state and symbolic approaches addressing different use cases.

In summary, our contribution is threefold. First, we introduce the first framework for model checking in zero knowledge, combining cryptographic principles with proof techniques for formal verification. Second, we present two schemes that realise this framework: an explicit-state algorithm for finite-state transition graphs and a symbolic algorithm for guarded command languages, whose thousands of proof obligations we fold into a single instance with a one-shot folding scheme. Third, we implement both schemes in a prototype and demonstrate the practical efficacy of our framework on LTL model checking tasks applied to standard communication and coordination protocols.

\section{Zero-Knowledge Model Checking}\label{sec:zkmc}

We address the problem of convincing an external verifier that a  state transition system
$\cal M$ satisfies a temporal specification $\phi$, denoted ${\cal M}\models \phi$, without revealing any information about
the initialisation or the transition structure of $\cal M$.

Our framework combines three complementary technologies.
First, we require a proof certificate witnessing $\mathcal{M} \models \phi$ and that can be verified in polynomial time. Second, we employ a commitment scheme that binds to $\mathcal{M}$ while keeping it hidden, ensuring that the system is uniquely identifiable through its cryptographic commitment. Third, we use a zero-knowledge proof scheme to convince an external verifier that (1) the proof certificate is valid and (2) it correctly corresponds to $\mathcal{M}$ and the specification $\phi$, without revealing any information about the system $\mathcal{M}$.

A {\em formal certification} scheme is an algorithm $\mbox{Cert}({\mathcal{M}}, \phi)$ that takes as input a {\em secret} encoding of the state transition system ${\mathcal{M}}$ and a {\em public} correctness specification $\phi$, and returns a {\em public} certificate $v_\phi$ that $\cal M$ satisfies $\phi$.
Crucially, we require the certification to be sound with respect to the model checking problem, in the sense that
\begin{equation}
    \mathcal{M} \not \models\phi  \implies  \mbox{Cert}(\mathcal{M}, \phi) \not\leadsto v_\phi,
\end{equation}
where $\mbox{Cert}(\mathcal{M}, \phi) \leadsto v_\phi$ denotes that $v_\phi$ is a valid output of $\mbox{Cert}(\mathcal{M},\allowbreak \phi)$.
We do not require the certification to be complete for the model checking problem,
a property that is desirable but not essential to implement our framework. Instead, we require that, given a certificate $v_\phi$, checking whether $\mbox{Cert}(\mathcal{M}, \phi) \overset{?}{\leadsto} v_\phi$---that is, whether the certificate is valid---is decidable in polynomial time.

A {\em cryptographic commitment} to $\mathcal{M}$ under a security parameter $\mathsf{sp}$ and randomness $r$ is denoted by $\mathrm{Com}(\mathsf{sp}, \mathcal{M}; r)$. To commit to $\mathcal{M}$, a party samples $r$ uniformly at random and computes the commitment $c_{\mathcal{M}} = \mathrm{Com}(\mathsf{sp}, \mathcal{M}; r)$. The commitment can later be opened by revealing $\mathcal{M}$ and $r$; validity is verified by checking whether $c_{\mathcal{M}} \overset{?}{=} \mathrm{Com}(\mathsf{sp}, \mathcal{M}; r)$.
We require the commitment scheme to satisfy the standard cryptographic properties of hiding and binding. Informally, hiding ensures that $c_{\mathcal{M}}$ reveals no information about $\mathcal{M}$, while binding guarantees that it is infeasible to produce a commitment $c_{\mathcal{M}}$ that can be opened to a different value
$\mathcal{M}' \neq \mathcal{M}$, thus preventing equivocation~\cite[p.~188]{KatzLindell2014}.

A {\em zero-knowledge proof} system for model checking is therefore a zero-knowledge proof system for the proof of knowledge $\mathcal{R}(w,u)$ below,
in the standard sense~\cite{DBLP:conf/crypto/BellareG92} with secret witness $w$ and public statement $u$:
\begin{multline}
    {\cal R} = \pok\{ \overbrace{({\mathcal{M}}, r)}^{w}; (\overbrace{v_\phi, c_{\cal M}}^{u}) : \\\mbox{Cert}({\mathcal{M}}, \phi) \overset{?}{\leadsto} v_\phi \land \mbox{Com}({\sf sp}, {\cal M}; r) \overset{?}{=} c_{\cal M}\}.
\end{multline}
In other words, given appropriate certification and commitment schemes, our objective is to convince an external verifier that we know a model $\mathcal{M}$ and randomness $r$ consistent with both the public certificate $v_\phi$ and the public commitment $c_{\cal M}$, thereby demonstrating that we possess a system $\mathcal{M}$ satisfying the public specification $\phi$.

\begin{figure}[t]
    \centering
    \begin{tikzpicture}[box/.style={rectangle,draw,minimum width=8mm,minimum height=7mm}]
        \def\vdist{5mm}
        \def\hdist{20mm}
        \def\vdistspec{15mm}
        \node[box] (prover) {$\cal P$};
        \node[box, above left=\vdist and \hdist of prover.center, anchor=center] (cert) {Cert};
        \node[box, below left=\vdist and \hdist of prover.center, anchor=center] (com) {Com};
        \node[box, right=1.25*\hdist of prover.center, anchor=center] (verifier) {$\cal V$};
        \node[below left=\vdist and \hdist of cert.center, anchor=center] (model) {$\cal M$};
        \node at ($(cert)!0.5!(com)$) (dummy) {};
        \node[above=\vdistspec, anchor=center] at ($(dummy)!0.5!(verifier)$) (spec) {$\phi$};
        \node at ($(verifier)+(0.5*\hdist,\vdist)$) (one) {1};
        \node at ($(verifier)+(0.5*\hdist,-\vdist)$) (zero) {0};

        \draw (model) -- ($(model.center)+(0.5*\hdist,0)$)
        ($(model.center)+(0.5*\hdist,0)$) -- ($(cert.center)-(0.5*\hdist,0)$)
        ($(cert.center)-(0.5*\hdist,0)$) edge[->] (cert)
        ($(model.center)+(0.5*\hdist,0)$) -- ($(com.center)-(0.5*\hdist,0)$)
        ($(com.center)-(0.5*\hdist,0)$) edge[->] (com);
        \draw (cert) -- ($(cert.center)+(0.5*\hdist,0)$) node[above] {$v_\phi$}
        ($(cert.center)+(0.5*\hdist,0)$)  --  ($(prover.center)-(0.5*\hdist,0)$)
        (com) -- ($(com.center)+(0.5*\hdist,0)$) node[below] {$c_{\cal M}$}
        ($(com.center)+(0.5*\hdist,0)$) -- ($(prover.center)-(0.5*\hdist,0)$)
        ($(prover.center)-(0.5*\hdist,0)$) edge[->] (prover);
        \draw ($(prover.north east)!0.5!(prover.south east)$) edge[->] node[above] {$v_\phi,c_{\cal M},  \pi$} ($(verifier.north west)!0.5!(verifier.south west)$);
        \draw (verifier) edge[->] (one);
        \draw (verifier) edge[->] (zero);
        \draw (spec) -- ($(dummy.center)+(0,\vdistspec)$)
        ($(dummy.center)+(0,\vdistspec)$) edge[->] (cert)
        (spec) -- ($(verifier.center)+(0,\vdistspec)$)
        ($(verifier.center)+(0,\vdistspec)$) edge[->] (verifier);

        \node[draw, rectangle, dashed, fit=(model)(prover)(cert)(com), inner sep=6pt] (secret) {};
        \node[below=1mm of secret, inner sep=0] {\scriptsize secret};
    \end{tikzpicture}
    \caption{Non-Interactive ZKMC Scheme}
    \label{fig:zkmc}
\end{figure}
As illustrated in \cref{fig:zkmc}, we propose a non-interactive scheme for zero-knowledge model checking, in which the prover computes a proof $\pi$ once, and the verifier checks it alongside the commitment $c_{\cal M}$ and the certificate $v_\phi$.

We assume a probabilistic polynomial-time adversary that may corrupt either party. The cryptographic guarantees of our framework follow the standard formulation of zero-knowledge proofs: completeness guarantees that an honest prover can always convince the verifier when the statement is true,
soundness ensures that the verifier detects any attempt by a malicious prover to establish a false statement except with negligible probability $\epsilon$, and zero knowledge against an honest-but-curious verifier with respect to the protocol transcript ensures that no information beyond the validity of the statement is revealed.

\begin{remark}[Soundness and Completeness]\label[remark]{remark:soundness}
  We stress that soundness and completeness for the certifier with respect to the model checking problem are distinct from those of the zero-knowledge proof system with respect to the commitment and certification itself, and must not be conflated. Under our definitions, a malicious prover $\mathcal{P}$ cannot, except with negligible probability $\epsilon$, convince a verifier $\mathcal{V}$ that $\mathcal{M} \models \phi$ when this is false, or when $\cal M$ does not correspond to its commitment:
\begin{multline}
{\cal M} \not\models \phi \lor  c_{\cal M} \neq  \mbox{Com}({\sf sp}, {\cal M}; r) \\\implies \bbbp\left[\mathsf{out}_{\cal V}\langle \mathcal{P}(w,u), \mathcal{V}(u) \rangle = 1 \right] \leq \epsilon
\end{multline}
where $\mathsf{out}_{\cal V}\langle \cdot \rangle$ denotes the final output of $\cal V$ upon interaction with $\cal P$.
Moreover, an honest prover can always convince the verifier when it possesses a valid certificate; a negative outcome, however, does not imply that $\mathcal{M} \not\models \phi$.
\end{remark}

\begin{remark}[Specification- and Certification-Induced Leakage]
    Our framework ensures that neither the commitment $c_{\cal M}$ nor the proof $\pi$ reveals any information about $\cal M$. However, selecting a secure specification $\phi$ and certificate $v_\phi$ remains the user's responsibility. In the extreme case where the public specification $\phi$ or the certificate $v_\phi$ uniquely determine the secret system $\mathcal{M}$, any party capable of verifying the proof could, in principle, reconstruct the system regardless of the protocol's cryptographic guarantees. Developing systematic specification and certification schemes that mitigate this risk constitutes an open problem, beyond the scope of this work.
\end{remark}

Side channels arising from the prover's computation, denial-of-service attacks, or attacks that combine our protocol with auxiliary channels (such as the timing of certificate generation, which may depend on $\mathcal{M}$) are addressed by standard countermeasures in the literature.

We instantiate our framework to LTL model checking in \Cref{sec:ltl} and present two alternative commitment and zero-knowledge schemes in \Cref{sec:explicit,sec:symbolic}.

\section{LTL Model Checking}\label{sec:ltl}

We instantiate our framework to model checking systems with discrete,
potentially infinite state spaces against LTL, more generally $\omega$-regular properties~\cite{DBLP:conf/podc/MannaP89}.
We leverage $\omega$-automata and proof certificates expressed as ranking functions to design a certifier,
thereby laying the foundations for our two alternative ZKMC algorithms in \Cref{sec:explicit,sec:symbolic}.

A nondeterministic state transition system $\cal M$ that executes over an infinite time horizon on a discrete state space $S$ is defined in terms of an initial condition $S_0 \subseteq S$ and a transition relation $T \subseteq S \times S$.
An execution of $\mathcal{M}$ is any infinite sequence of states $(s_i \in S)_{i \in \bbbn}$ such that $s_0 \in S_0$ and $(s_i, s_{i+1}) \in T$ for all $i \in \bbbn$.
We verify temporal logic specifications over a finite set of atomic propositions $\Pi$, and denote by $\Sigma = 2^\Pi$ the alphabet of all possible observations over states. For a state $s$, we write $\llangle s \rrangle \in \Sigma$ for the set of propositions that hold in that state. Each execution $(s_i)_{i \in \bbbn}$ then induces a trace defined as the infinite sequence $(\llangle s_i \rrangle)_{i \in \bbbn}$, representing the corresponding observable behaviour. Finally, we define the language $L_{\mathcal{M}} \subseteq \Sigma^\omega$ of $\mathcal{M}$ as the set of all traces induced by its executions.

\begin{figure*}[h]
    \centering
    \def\bwidth{14mm}
    \def\swidth{9mm}
    \def\hdist{4mm}
    \def\bdist{6mm}
    \def\vdist{14mm}
    \def\segsize{2mm}
    \def\splitone{0.63}
    \def\splittwo{0.37}
    \setlength{\tabcolsep}{0pt}
\begin{tabular}{cc}
\parbox[c]{\splitone\linewidth}{\centering\begin{tikzpicture}[box/.style={rectangle,draw,minimum width=\bwidth,minimum height=7mm, inner sep=2pt},
    event/.style={midway, sloped, above=0.5pt, inner sep=1pt, fill=white}, font=\tiny]
    \node[box, minimum width=\swidth] (closed) {\tt Closed};

    \node[box, above right=\vdist and \bdist of closed, align=left] (w10-0) {\makebox[\bwidth]{\tt Wait}\\{\tt delay} = $2^{8}-1$\\{\tt done} = $0$};
    \node[box, right=\bdist  of closed, align=left] (w6-0) {\makebox[\bwidth]{\tt Wait}\\{\tt delay} = $2^{6}-1$\\{\tt done} = $0$};
    \node[box, below right=\vdist and \bdist of closed, align=left] (w0-0) {\makebox[\bwidth]{\tt Wait}\\{\tt delay} = $0$\\{\tt done} = $0$};

    \node[box, right=\hdist of w10-0, align=left] (w10-1) {\makebox[\bwidth]{\tt Wait}\\{\tt delay} = $2^{8}-1$\\{\tt done} = $1$};
    \node[box, below=0.45*\vdist of w10-1, align=left] (w7-1) {\makebox[\bwidth]{\tt Wait}\\{\tt delay} = $2^{7}-1$\\{\tt done} = $1$};
    \node[box, right=\hdist of w0-0, align=left] (w0-1) {\makebox[\bwidth]{\tt Wait}\\{\tt delay} = $0$\\{\tt done} = $1$};

    \node[box, right=\hdist of w10-1, align=left] (w10-2) {\makebox[\bwidth]{\tt Wait}\\{\tt delay} = $2^{8}-1$\\{\tt done} = $2$};
    \node[box, right=\hdist of w0-1, align=left] (w0-2) {\makebox[\bwidth]{\tt Wait}\\{\tt delay} = $0$\\{\tt done} = $2$};

    \node[box, right=\hdist of w10-2, align=left] (w10-3) {\makebox[\bwidth]{\tt Wait}\\{\tt delay} = $2^{8}-1$\\{\tt done} = $3$};
    \node[box, right=\hdist of w0-2, align=left] (w0-3) {\makebox[\bwidth]{\tt Wait}\\{\tt delay} = $0$\\{\tt done} = $3$};

    \node[box, below right=\vdist and \bdist of w10-3, minimum width=\swidth] (success) {\tt Established};
    \node[box, below=3mm of w0-2, minimum width=\bwidth, align=left] (exhausted) {\makebox[\bwidth]{\tt Exhausted}\\{\tt done} = $3$};

    \draw[-] ($(w10-0.south)$) -- ($(w10-0.south)-(0,\segsize)$);
    \draw[dashed] ($(w10-0.south)-(0,\segsize)$) -- ($(w6-0.north)+(0,\segsize)$);
    \draw[->] ($(w6-0.north)+(0,\segsize)$) -- ($(w6-0.north)$);

    \draw[-] ($(w6-0.south)$) -- ($(w6-0.south)-(0,\segsize)$);
    \draw[dashed] ($(w6-0.south)-(0,\segsize)$) -- ($(w0-0.north)+(0,\segsize)$);
    \draw[->] ($(w0-0.north)+(0,\segsize)$) -- ($(w0-0.north)$);

    \draw[->, dashed] (w10-1) -- (w7-1);

    \draw[-] ($(w7-1.south)$) -- ($(w7-1.south)-(0,\segsize)$);
    \draw[dashed] ($(w7-1.south)-(0,\segsize)$) -- ($(w0-1.north)+(0,\segsize)$);
    \draw[->] ($(w0-1.north)+(0,\segsize)$) -- ($(w0-1.north)$);

    \draw[-] ($(w10-2.south)$) -- ($(w10-2.south)-(0,\segsize)$);
    \draw[dashed] ($(w10-2.south)-(0,\segsize)$) -- ($(w0-2.north)+(0,\segsize)$);
    \draw[->] ($(w0-2.north)+(0,\segsize)$) -- ($(w0-2.north)$);

    \draw[-] ($(w10-3.south)$) -- ($(w10-3.south)-(0,\segsize)$);
    \draw[dashed] ($(w10-3.south)-(0,\segsize)$) -- ($(w0-3.north)+(0,\segsize)$);
    \draw[->] ($(w0-3.north)+(0,\segsize)$) -- ($(w0-3.north)$);
    \draw[-] ($(w0-3.east)$) -- ($(w0-3.east)+(\segsize,0)$);
    \draw[-] ($(w0-3.east)+(\segsize,0)$) -- ($(w10-3.east)+(\segsize,0)$);
    \draw[->] ($(w10-3.east)+(\segsize,0)$) -- ($(w10-3.east)$);

    \draw[->] (closed) -- (w0-0);
    \draw[->] (w0-0) -- (w0-1);
    \draw[->] (w0-1) -- (w0-2);
    \draw[dashed] (closed) -- ($(closed)+(9mm,-4mm)$);
    \draw[dashed] (closed) -- ($(closed)+(8mm,-6mm)$);
    \draw[dashed] (w0-0) -- ($(w0-0)+(11mm,+10mm)$);
    \draw[dashed] (w0-0) -- ($(w0-0)+(12mm,+6mm)$);
    \draw[dashed] (w0-1) -- ($(w0-1)+(11mm,+10mm)$);
    \draw[dashed] (w0-1) -- ($(w0-1)+(12mm,+6mm)$);

    \draw[->] (w10-3) -- node[event] {{\tt synack}} (success);
    \draw[->] (w0-3) -- node[event]{{\tt synack}} (success);
    \draw[->] (w10-2) -- node[event] {{\tt synack}} (success);
    \draw[->] (w0-2) -- node[event] {{\tt synack}} (success);
    \draw[->] (w10-1) -- node[event] {{\tt synack}} (success);
    \draw[->] (w7-1) -- node[event] {{\tt synack}} (success);
    \draw[->] (w0-1) -- node[event] {{\tt synack}} (success);
     \draw[->] ($(w10-0.south east)+(0,2pt)$) -- node[event] {{\tt synack}} (success);
    \draw[->] ($(w0-0.north east)-(0,2pt)$) -- node[event] {{\tt synack}} (success);
    \draw[->] (w6-0) -- node[event] {{\tt synack}} (success);
    \draw[->] (closed) -- node[event] {{\tt syn}} (w6-0);
    \draw[->] (w0-1) -- node[event] {\tt syn}(w10-2);
    \draw[->] (w0-0) -- node[event] {\tt syn} (w7-1);
    \draw[->] (w0-2) -- (exhausted);
    \draw[->] (exhausted) edge[loop right, out=10, in=-10, looseness=8] (exhausted);
    \draw[->] (closed) edge[loop above] (closed);
    \draw[->] ($(closed.west)-(3mm, 0)$) -- (closed);
    \draw[->] (success) edge[loop above] node[event] {{\tt ack}} (success);

    \end{tikzpicture}}
    &
    \parbox[c]{\splittwo\linewidth}{\centering \begin{tikzpicture}[>=Stealth,node distance=2.5cm, initial text=]
    \node[state, initial] (q0) {$q_0$};
    \node[state, right of=q0] (q1) {$q_1$};

    \draw[->, >=Stealth] (q0) edge[loop above] node{\text{true}} (q0);
    \draw[->, >=Stealth] (q0) edge[above] node{\tt Wait} (q1);
    \draw[->, >=Stealth] (q1) edge[loop above, double] node{\tt Wait}(q1);
    \end{tikzpicture}}
    \\[0.5cm]
    \parbox[c]{\splitone\linewidth}{\centering (a)} &
    \parbox[c]{\splittwo\linewidth}{\centering (b)}
\end{tabular}

    \caption{(a) 3-Way Handshake with Exponential Backoff and (b) Nondeterministic B\"uchi Automaton for $\finally\globally\,\tt{Wait}$}
    \label{fig:backoff}
\end{figure*}

\begin{example}[3-Way Handshake with Exponential Backoff]
    The 3-way handshake protocol with exponential backoff is a mechanism for establishing a persistent connection between two parties. It is implemented in numerous communication protocols, such as TCP \cite[p.~30]{rfc793}, and provides a paradigmatic example of a model checking problem. \Cref{fig:backoff}a illustrates the protocol from the initiator's perspective. The initiator begins by sending a {\tt syn} message and waits for the listener to respond with a {\tt synack}. Upon receiving this response, the connection is established, and the initiator responds with {\tt ack} while transmitting data. The waiting time is sampled randomly (modelled nondeterministically), initially from the range $[0, 2^6 - 1]$. If no {\tt synack} is received, the {\tt syn} message is retransmitted and the waiting time is resampled from a range of double size. This process is repeated for at most three attempts in this example. In our model, the system state is determined by four mutually exclusive Boolean variables {\tt Closed}, {\tt Wait}, {\tt Established}, and {\tt Exhausted}, an 8-bit unsigned integer {\tt delay}, and a 2-bit unsigned integer {\tt done}. The variables {\tt delay} and {\tt done} are relevant only when the system is in state {\tt Wait}.
\end{example}

An LTL formula $\phi$ over the atomic predicates $\Pi$ extends propositional logic with temporal operators: the next operator $\lnext \phi$ indicates that $\phi$ holds at the next step; the eventually (finally) operator $\finally \phi$ indicates that $\phi$ holds at some time in the future; the always (globally) operator $\globally \phi$ indicates that $\phi$ holds at all future times; the until operator $\phi_1 \until \phi_2$ indicates that $\phi_1$ holds continually until some future time where $\phi_2$ holds~\cite{DBLP:conf/focs/Pnueli77}.
We define the language $L_\phi \subseteq \Sigma^\omega$ of a specification $\phi$ as the set of all traces $(\sigma_i)_{i \in \bbbn}$ that satisfy $\phi$.
The LTL model checking problem $\mathcal{M} \models \phi$ asks whether all traces of $\mathcal{M}$ satisfy $\phi$, which is the language inclusion problem $L_{\mathcal{M}} \subseteq L_\phi$.

We rely on the standard result that LTL is a special case of $\omega$-regular specifications, and that every $\omega$-regular specification can be recognised by a nondeterministic B\"uchi automaton $\mathcal{A}$ accepting the same language. Such an automaton is defined by a finite set of states $Q$, with the initial states $Q_0 \subseteq Q$, and a transition relation $\delta \subseteq Q \times \Sigma \times Q$, with the set of fair (or accepting) transitions $F \subseteq \delta$.
An input trace $(\sigma_i)_{i \in \bbbn}$ is accepted by $\cal A$ if and only if there exists an infinite sequence of states $(q_i \in Q)_{i \in \bbbn}$ such that $q_0 \in Q_0$ and $(q_i, \sigma_i, q_{i+1}) \in \delta$ for all $i \in \bbbn$, and infinitely many fair transitions are visited, i.e., $\sum_{i=0}^\infty {\bf 1}_F(q_i, \sigma_i, q_{i+1}) = \infty$\footnote{For a set $A$, $\mathbf{1}_A$ denotes its indicator function.}.
The language $L_{\mathcal{A}}$ of the automaton $\mathcal{A}$ is the set of all traces it accepts. For every LTL formula $\phi$, one can construct (using appropriate algorithms~\cite{DBLP:conf/cav/Duret-LutzRCRAS22}) a nondeterministic B\"uchi automaton $\mathcal{A}_\phi$ such that $L_{\mathcal{A}_\phi} = L_\phi$.

We adopt the standard automata-theoretic approach to LTL model checking~\cite{DBLP:conf/lics/VardiW86}. We first construct a nondeterministic B\"uchi automaton $\mathcal{A}_{\lnot \phi}$ for the complementary specification $\lnot \phi$, and then reduce model checking to the language emptiness problem $L_{\mathcal{M}} \cap L_{\mathcal{A}_{\lnot \phi}} = \emptyset$. Since $L_{\mathcal{A}_{\lnot \phi}} = \Sigma^\omega \setminus L_{\mathcal{A}_\phi}$, this is equivalent to the language inclusion $L_{\mathcal{M}} \subseteq L_{\mathcal{A}_\phi}$, which in turn coincides with $L_{\mathcal{M}} \subseteq L_\phi$, that is, precisely the original model checking question $\mathcal{M} \models \phi$.

Our goal is to convince an external verifier that $\mathcal{M} \models \phi$ and, prior to executing our zero knowledge protocol,
we require an appropriate certificate that no trace of $\cal M$ is accepted by $\mathcal{A}_{\lnot \phi}$.
This amounts to certifying that no execution of the synchronous composition of $\cal M$ and $\mathcal{A}_{\lnot \phi}$ is fair, which establishes that its language  $L_{\mathcal{M}} \cap L_{\mathcal{A}_{\lnot \phi}}$ is empty.
We define a fair execution of the synchronous composition of $\mathcal{M}$ and $\mathcal{A}_{\lnot \phi}$ as any infinite sequence $((s_i, q_i) \in S \times Q)_{i \in \bbbn}$ over the composed state space such that (i) $s_0 \in S_0$ and $q_0 \in Q_0$,
(ii) $(s_i, s_{i+1}) \in T$ and $(q_i, \llangle s_i \rrangle, q_{i+1}) \in \delta$ for all $i \in \bbbn$, and (iii) fair transitions are visited infinitely often, i.e., $\sum_{i=0}^\infty {\bf 1}_{F}(q_i, \llangle s_i \rrangle,  q_{i+1}) = \infty$.
Our objective is to show that no fair execution exists, which constitutes a fair termination problem~\cite{DBLP:conf/popl/CookGPRV07,Vardi91,DBLP:journals/iandc/GrumbergFR85,DBLP:conf/nips/GiacobbeKPT24}.

Our certification scheme relies on a ranking function
that assigns to each composed state an upper bound on the number of fair transitions that can possibly occur along any execution starting from that state~\cite[Theorem 4.4]{Vardi91}.

\begin{theorem}\label{thm:ranking} Let $\Sigma$ be a finite alphabet. Let $\cal M$ be a state transition system with state space $S$, initial condition $S_0 \subseteq S$, transition relation $T \subseteq S \times S$, and observation function $\llangle \cdot \rrangle \colon S \to \Sigma$. Let $\cal A$ be a nondeterministic B\"uchi automaton with state space $Q$, initial states $Q_0 \subseteq Q$, transition relation $\delta \subseteq Q \times \Sigma \times Q$, and fair transitions $F \subseteq \delta$.
Suppose there exists a function $V \colon S \times Q \to \mathbb{N} \cup \{ +\infty \}$ such that, for every $s,s' \in S$ and $q,q' \in Q$, the following three conditions hold:
\begin{align}
s \in S_0 &\implies \bigl(q \in Q_0 \implies V(s,q) \neq \infty \bigr)\\
(s,s') \in T &\implies
  \begin{multlined}[t]
    \bigl( V(s,q) \neq \infty \wedge (q,\llangle s\rrangle,q') \in \delta \\
    \implies V(s,q) \geq V(s',q')\bigr)
  \end{multlined} \centeredtag\\
(s,s') \in T &\implies
  \begin{multlined}[t]
    \bigl( V(s,q) \neq \infty \wedge (q,\llangle s\rrangle,q') \in F \\
    \implies V(s,q) > V(s',q') \bigr)
  \end{multlined} \centeredtag
  \\[-2ex]
\underbrace{\phantom{\smash{(s,s') \in T}}}_{\text{secret}}
    &
\phantom{\smash \implies\;} \underbrace{\phantom{\smash{ \begin{multlined}[t]
    V(s,q) \neq \infty \wedge (q,\llangle s\rrangle,q') \in F \\
    \implies V(s,q) > V(s',q')
  \end{multlined}}}}_{\text{public}}
    \notag
\end{align}
Then, $L_{\cal M} \cap L_{\cal A} = \emptyset$. We call $V$ a ranking function for the fair termination of the synchronous composition of $\cal M$ and $\cal A$.
\end{theorem}

Intuitively, the first condition ensures that all initial composed states are assigned a finite bound (base case),
the second condition enforces monotonicity of the ranking function along all transitions (inductive step), and
the third condition guarantees strict progress whenever a fair transition is taken, thereby ruling out infinite fair executions (fair step).

\begin{corollary}\label[corollary]{cor:cert}
Let $\cal M$ be a state transition system. Let $\phi$ be an LTL formula and ${\cal A}_{\lnot \phi}$ be a nondeterministic B\"uchi automaton recognising $L_{\lnot \phi}$. If there exists a ranking function for the fair termination of the synchronous composition of $\cal M$ and ${\cal A}_{\lnot \phi}$, then ${\cal M} \models \phi$.
\end{corollary}

\begin{example}[Wait, but not Forever]
An important correctness requirement for the protocol in \cref{fig:backoff}a is that it must never get stuck when the listener is unreachable; in other words, it should never wait forever. This liveness property is naturally expressed in LTL by the formula $\lnot \finally \globally\,{\tt Wait}$, which rules out executions in which the system eventually remains in the waiting state indefinitely. \Cref{fig:backoff}b depicts a B\"uchi automaton for the complementary property $\finally \globally\,{\tt Wait}$. The automaton has two states and three transitions in total, one of which is fair (the self-loop on $q_1$, shown with a double mark). Our criteria require a ranking function
$V$ over the combined state variables of the system and the automaton to take finite values on all reachable states. In automaton state
$q_0$, all transitions are permitted, and $V$ is required to be non-increasing along every transition. In contrast, state $q_1$ restricts the analysis
to only the transitions in {\tt Wait} and requires $V$ to decrease strictly, thereby certifying that the system can remain in {\tt Wait} only for a finite duration. A suitable ranking function is the following piecewise-defined function:
\begin{align*}
    &V(\cdots,q_1) =\\
    &\begin{cases}
        3 \cdot 2^8 &\text{if }\tt{Closed} \\
        2^9 + 1 - 2^8\!\cdot{\tt done} + {\tt delay} & \text{if }\tt{Wait} \land {\tt done} \leq 2\\
        0&\text{if }{\tt Established}\\
        0&\text{if } {\tt Exhausted}\\
        +\infty &\text{if }{\tt Wait} \land {\tt done} = 3
    \end{cases}\allowdisplaybreaks[4]\\
    &V(\cdots,q_0) = \begin{cases}
        +\infty &\text{if } {\tt Wait} \land {\tt done} = 3\\
        3 \cdot 2^8 &\text{otherwise}
    \end{cases}
\end{align*}
\end{example}

Overall, \Cref{cor:cert} forms the basis for a certification scheme for LTL model checking, in which the automaton and the ranking function jointly constitute the public certificate $v_\phi = ({\cal A}_{\lnot \phi}, V)$.  The existence of such a ranking function is therefore a sufficient condition for a positive answer to the model checking problem. We remark that it is not a necessary condition in general; this affects only the completeness of our approach, not its soundness.

Our framework thus requires a certifier that is checkable in polynomial time and admits a commitment and zero-knowledge proof scheme. We present two such strategies: an explicit-state approach that enumerates the state space (\Cref{sec:explicit}), and a symbolic approach for transition systems expressed as linear guarded commands (\Cref{sec:symbolic}).

\section{Explicit-State ZKMC}\label{sec:explicit}

We assume that the state space of $\mathcal{M}$ is finite and that explicit enumeration is feasible, in the spirit of explicit-state model checking~\cite{DBLP:reference/mc/Holzmann18}.
Our scheme proceeds in two phases. First, both parties enumerate all hypothetical states and transitions that can possibly violate the obligations of \cref{sec:ltl} using public information. Then, the prover convinces the verifier  that none of the elements in these batches belong to the system without revealing any secret information.

The public information comprises the global state space $S$, the automaton $\mathcal{A}_{\lnot \phi}$ (its states $Q$, initial states $Q_0$, transitions $\delta$, and fair transitions $F$), and the ranking function $V$.
The secret information, known only to the prover, is the system's initial states $S_0$ and transition relation $T$.

Our certification scheme proceeds by proving the three proof obligations of \cref{thm:ranking} {\em by contraposition}. Each obligation has the form of an implication with a secret premise and a public consequent. We therefore negate the public consequent to identify batches of states that violate it, and then prove that none of these states satisfy the secret premises given by $S_0$ and $T$.

\begin{figure*}[h]
\centering
\def\constrwidth{\textwidth}
\begin{minipage}{\constrwidth}
\centering\scriptsize
\setlength{\tabcolsep}{0pt}
\begin{tabular}{|p{\constrwidth}|}
    \hline
    \textbf{Prover Input:} Secret polynomials $p_{S_0}, p_T$, randomness $r_{S_0}, r_T$ \\
    \textbf{Common Input:} Security parameter $1^\lambda$, degree bound $t$, certificate $v_\phi = ({\cal A}_{\lnot \phi}, V)$
    \\\hline
    \centering\textbf{Setup Phase} (both parties) \tabularnewline\hline
    $\mathsf{sp} \gets \mathsf{kzg.Setup}(1^\lambda, t)$ \\
    \hline
\end{tabular}
\begin{tabular}{|p{0.5\constrwidth} | p{0.5\constrwidth}|}
    \centering\textbf{Prover} & \centering\textbf{Verifier} \tabularnewline\hline
    \begin{minipage}[t]{0.5\constrwidth}
    Compute $E_{\sf init}, E_{\sf step}, E_{\sf fair}$ from $v_\phi$\\
    $c_{S_0} \gets \mathsf{kzg.Commit}(\mathsf{sp}, p_{S_0}; r_{S_0})$\\
    $c_T \gets \mathsf{kzg.Commit}(\mathsf{sp}, p_T; r_T)$\\
    {\bf send} $c_{\cal M} = (c_{S_0}, c_T)$\\[-2pt]
    \noindent\makebox[\linewidth]{\dotfill}
    $\pi_{S_0} \gets \mathsf{kzg.Prove}(\mathsf{sp}, p_{S_0}, c_{S_0}, r_{S_0}, E_{\sf init}, 0)$\\
    $\pi_T \gets \mathsf{kzg.Prove}(\mathsf{sp}, p_T, c_T, r_T, E_{\sf step} \cup E_{\sf fair}, 0)$\\
    {\bf send} $\pi = (\pi_{S_0}, \pi_T)$
    \end{minipage}&
    \begin{minipage}[t]{0.5\constrwidth}
    Compute $E_{\sf init}, E_{\sf step}, E_{\sf fair}$ from $v_\phi$\\
    {\bf receive} $c_{\cal M} = (c_{S_0}, c_T)$ \\[-2pt]
    \noindent\makebox[\linewidth]{\dotfill}
    {\bf receive} $\pi = (\pi_{S_0}, \pi_T)$\\
    $\mathsf{kzg.Verify}(\mathsf{sp}, c_{S_0}, E_{\sf init}, 0, \pi_{S_0}) \stackrel{?}{=} 1$\\
    $\mathsf{kzg.Verify}(\mathsf{sp}, c_T, E_{\sf step} \cup E_{\sf fair}, 0, \pi_T) \stackrel{?}{=} 1$\\
    \textbf{accept}
    \end{minipage}
    \tabularnewline\hline
\end{tabular}
\end{minipage}
\caption{Construction of our Explicit-State ZKMC Scheme}
\label{fig:ZKMC-E}
\end{figure*}

Both parties enumerate the following three batches of states, which are the hypothetical states and transitions violating the consequence of our obligations:
\begin{align}
    B_{\sf init} &= \{ s \!\mid\! \exists q \mathpunct. q \in Q_0 \land V(s,q) = \infty \}\allowdisplaybreaks\\
    B_{\sf step} &= \{ (s,s') \!\mid\! \exists q,q'\mathpunct.\! V(s,q) \neq \infty \land (q, \llangle s\rrangle, q') \in \delta \notag\\
    &\hspace{0.46\linewidth}\land V(s,q) < V(s', q') \}\allowdisplaybreaks\\
    B_{\sf fair} &= \{ (s,s') \!\mid\! \exists q,q'\mathpunct.\! V(s,q) \neq \infty \land (q, \llangle s\rrangle, q') \in F  \notag\\
    &\hspace{0.46\linewidth}\land V(s,q) \leq V(s', q') \}
\end{align}
Our prover must then convince the verifier that
\begin{equation}\label{eq:explicit-cert}
    S_0 \cap B_{\sf init} = \emptyset,\qquad
    T \cap B_{\sf step} = \emptyset, \qquad
    T \cap B_{\sf fair}  = \emptyset,
\end{equation}
without revealing the secret $S_0$ and $T$ while committing to them, which is a set membership task. Crucially, $B_{\sf init}$,
$B_{\sf step}$, and $B_{\sf fair}$
are fully determined by public information.

Polynomial commitment schemes are a common candidate to realise set membership tasks, and we use the KZG scheme with Pedersen-style randomisation and batch evaluation~\cite{KateZG10}. We first embed states and transitions into a field, and then reduce the disjointness obligations to the vanishing of two committed polynomials.

We associate the state space $S$ and the transition space $S \times S$ with two public embeddings $\mbox{e}_1 \colon S \to \bbbf$ and $\mbox{e}_2 \colon S \times S \to \bbbf$ into an appropriate field $\bbbf$. Both embeddings are fixed, public, and injective, allowing both parties to independently compute the evaluation points for $B_{\sf init}$, $B_{\sf step}$ and $B_{\sf fair}$.

Our prover computes two secret polynomials on $\bbbf$: a polynomial $p_{S_0}$ derived from the set of initial states, and a polynomial $p_T$ derived from the transition relation of the system. These polynomials satisfy the following properties:
\begin{align}
    \forall s \in S &\colon \mathbf{1}_{S_0}(s) = p_{S_0} \circ \mbox{e}_1(s)\\
    \forall (s,s') \in S^2 &\colon \mathbf{1}_T(s,s') = p_T \circ \mbox{e}_2(s,s')
\end{align}
where, without loss of generality, we assume that the underlying field $\mathbb{F}$ contains the elements $0$ and $1$.

The prover and the verifier operate directly over the field $\bbbf$.
In particular, they construct the public encodings
$E_{\sf init} = \mbox{e}_1[ B_{\sf init}]$, $E_{\sf step} = \mbox{e}_2[ B_{\sf step}]$, and $E_{\sf fair} = \mbox{e}_2[ B_{\sf fair}]$,
where $f[A]$ indicates the image of a set $A$ through $f$. The prover's objective is then to convince the verifier, without revealing the secret polynomials $p_{S_0}$ and $p_T$, that the following conditions hold:
\begin{align}
    &\forall e \in E_{\sf init} \colon p_{S_0}(e) = 0 \label{eq:vanish1}\\
    &\forall e \in E_{\sf step} \cup E_{\sf fair}  \colon p_{T}(e) = 0\label{eq:vanish2}
\end{align}
In other words, checking that a public batch $B$ of states or transitions has empty intersection with its secret counterpart is equivalent to proving that a certain polynomial $p$ evaluates to zero over the corresponding evaluation points $E$.

From KZG, we only need the special case in which the committed polynomial \emph{vanishes} on a public evaluation set, captured by the relation
\begin{equation*}
    R_{\mathsf{eval}}=\pok\begin{Bmatrix}
        (p, r_p); (c_p, E) \colon \\
        c_p = \mathsf{Com}(\mathsf{sp}, p; r_p) \land \\
        \forall e \in E,\, p(e) = 0
    \end{Bmatrix}
\end{equation*}
Our scheme uses a prover $\mathsf{kzg.Prove}(\mathsf{sp}, p, c_p, r_p, E, 0) \to \pi$ that produces a proof $\pi$ for $R_{\mathsf{eval}}$, and a verifier $\mathsf{kzg.Verify}(\mathsf{sp}, c_p, E, 0, \pi) \to \{0,1\}$ that accepts if and only if $\pi$ is valid---that is, the polynomial committed in $c_p$ vanishes on every point of $E$.
The batch zero-check can be viewed as a special case of standard KZG batch opening~\cite{KateZG10}, where all evaluation values are zero, allowing the proof to be expressed solely via divisibility by the vanishing polynomial.

KZG requires a trusted setup that produces a structured reference string of powers of a secret $\tau$ in the bilinear group; we assume this setup is executed honestly (e.g., via a multi-party powers-of-tau ceremony) and treat the resulting common reference string as part of the public parameters $\mathsf{sp}$. Soundness rests on the $q$-strong Diffie--Hellman assumption over bilinear groups~\cite{KateZG10}.

The KZG scheme satisfies the standard properties of completeness, polynomial binding, evaluation binding, and polynomial hiding; used as a zero-knowledge proof, the scheme additionally provides knowledge soundness and zero knowledge in the sense of \Cref{sec:zkmc}.

We present our construction in \cref{fig:ZKMC-E}, which directly applies polynomial commitments to establish that $\mathcal{M}\models\phi$.
It implements the following proof of knowledge:
\begin{equation*}
    {\cal R}_{\sf e}=\pok\begin{Bmatrix}
        (p_{S_0}, p_T, r_{S_0}, r_T); (c_{S_0}, c_T, E_{\sf init}, E_{\sf step}, E_{\sf fair}) \colon \\
        \cref{eq:vanish1} \land \cref{eq:vanish2} \land \\
        \mathsf{Com}(\mathsf{sp}, p_{S_0}; r_{S_0}) = c_{S_0} \land \\
        \mathsf{Com}(\mathsf{sp}, p_T; r_T) = c_T
    \end{Bmatrix}
\end{equation*}
This construction composes several zero-knowledge sub-proofs, each executed with independent randomness, and inherits its security properties from the polynomial commitment scheme.
Soundness follows from binding and evaluation-binding: a cheating prover cannot produce accepting proofs unless the committed polynomials actually vanish on the evaluation sets, which by construction implies $S_0 \cap B_{\sf init} = \emptyset$ and $T \cap (B_{\sf step} \cup B_{\sf fair}) = \emptyset$. Zero knowledge follows from polynomial hiding and from the fact that all proof elements are masked by randomness.
Completeness is immediate from completeness of the underlying scheme.

\paragraph{Complexity}
The cost of our explicit-state scheme is dominated by the polynomial
$p_T$, whose degree is $\mathcal{O}(|S|^2)$ because its evaluation
domain ranges over all state pairs in $S \times S$. Committing to $p_T$
and generating the batch evaluation proof therefore take time
$\mathcal{O}(|S|^2)$, and the total batch size
$|E_{\sf init}| + |E_{\sf step}| + |E_{\sf fair}|$ grows accordingly.
The certificate itself, however, remains succinct: the batch evaluation
feature of the KZG scheme produces a proof consisting of just two group
elements, regardless of the size of the state space.

\section{Symbolic ZKMC}\label{sec:symbolic}

We allow the state space of $\mathcal{M}$ to be arbitrarily large, potentially countably infinite, and adopt a symbolic approach to certification, in the spirit of symbolic model checking~\cite{DBLP:conf/lics/BurchCMDH90}.
We require the state transition system to be expressed over integer variables, via guarded commands with linear constraints.
Under this setting, we present a certifier that leverages Farkas' lemma to establish the validity of a ranking function as in \cref{sec:ltl}, and a zero-knowledge proof scheme based on range proofs and matrix multiplication proofs.

We consider systems specified programmatically using guarded command languages~\cite{Lamport83, AlurH96}, and restrict our attention to systems over integer variables and coefficients, whose initialisation and update commands are linear:
\begin{align}
    S_0 &= \{x \mid A_0 x \leq b_0 \}
    \\
    T &= \{ (x,x') \mid \underbrace{A_1\begin{bmatrix}
         x\\x'
     \end{bmatrix} \leq b_1}_{\mathclap{\text{guarded update}_1}} \lor \dots \lor
     \underbrace{A_n\begin{bmatrix}
         x\\x'
     \end{bmatrix} \leq b_n}_{\mathclap{\text{guarded update}_n}}  \}
\end{align}
We denote by $x$ the variables in the current state and by $x'$ those in the next state.
The initialisation is given by a single system of linear inequalities $A_0 x \leq b_0$, whereas the transition relation is the union of $n$ guarded update rules, each specified as a system of linear inequalities $A_i [x~x']^{\sf T} \leq b_i$ over both the current and the next state, for $i = 1,\dots,n$.

We represent automata explicitly and assume that each atomic predicate in $\Pi$ corresponds to a linear inequality over the state variables $x$. Consequently, every valuation $\sigma \in \Sigma$ of atomic predicates induces a system of linear inequalities:
\begin{equation}
    \sigma = \llangle x \rrangle \iff \underbrace{P^{(\sigma)} x \leq r^{(\sigma)}}_{\mathclap{{\text{atomic predicates}}}}
\end{equation}
Since all variables are integer, negated non-strict inequalities can be encoded as non-strict inequalities as well, by applying an appropriate offset of~$1$.

We allow ranking functions to be specified separately for each automaton state as piecewise linear functions~\cite{DBLP:conf/tacas/ColonS01,DBLP:conf/tacas/UrbanGK16,DBLP:conf/vmcai/PodelskiR04} defined over polyhedral regions:
\begin{equation}
    V(x,q) = \begin{cases}
         w^{(q)}_1 x + u^{(q)}_1 & \text{if } C_1^{(q)} x \leq d_1^{(q)}\\
        \quad\vdots\\
        w^{(q)}_m x + u^{(q)}_m & \text{if } C_m^{(q)} x \leq d_m^{(q)}\\
        +\infty & \text{if } E_{1}^{(q)} x \leq f_{1}^{(q)}\\
        \quad\vdots\\
        +\infty & \text{if } E_l^{(q)} x \leq f_l^{(q)}\\
    \end{cases}
\end{equation}
To ensure soundness, we require the case constraints to define a total function with a non-negative codomain. Accordingly, the verifier must check that the case constraints are (i) pairwise disjoint, (ii) jointly cover the entire state space, and (iii) ensure that the function takes non-negative values on its entire domain $S \times Q$. Since the function is public, all of these checks can be carried out using standard constraint-solving or SMT solving techniques~\cite{DBLP:series/faia/BarrettSST09,DBLP:series/txtcs/KroeningS16,DBLP:journals/cacm/MouraB11}.

The conditions of \cref{thm:ranking} give rise to proof obligations that can be expressed as Boolean combinations of linear inequalities.
In particular, the initiation condition requires the ranking function to be finite on all initial states; equivalently, for every initial automaton state $q \in Q_0$ and every infinite case \(k = 1,\dots,l\), no initial state may satisfy the case constraint:
\begin{equation}\label{eq:cond-matrix-init}
    \underbrace{A_0 x \leq b_0}_{x\in S_0} \implies E^{(q)}_k x \not\leq f^{(q)}_k
\end{equation}

The condition of the inductive step requires that finiteness of the ranking function be preserved across transitions. Thus, for every guarded update $i = 1, \dots, n$, every automaton transition $(q,\sigma,q')\in\delta$, and every finite case $j = 1,\dots,m$, no successor state may satisfy an infinite case $k = 1,\dots,l$:
\begin{equation}\label{eq:cond-matrix-finite}
    \underbrace{A_i \begin{bmatrix}
        x\\x'
    \end{bmatrix} \leq
    b_i}_{(x,x') \in T}
     \implies
    \underbrace{\begin{bmatrix}
        P^{(\sigma)}  \\
        C_j^{(q)}
    \end{bmatrix} x \leq \begin{bmatrix}
         r^{(\sigma)}\\
         d_j^{(q)}
    \end{bmatrix}}_{\mathclap{\llangle x \rrangle = \sigma \land V(x,q) \neq \infty}}
     \implies E^{(q')}_k x' \not\leq f^{(q')}_k
\end{equation}

Finally, for every guarded update $i = 1,\dots,n$, every automaton transition $(q,\sigma,q') \in \delta$, and every pair of finite cases $j,k = 1,\dots,m$, we require the following ranking condition:
\begin{multline}\label{eq:cond-matrix-step}
    \overbrace{A_i \begin{bmatrix}
        x\\x'
    \end{bmatrix} \leq
    b_i}^{(x,x') \in T}
     \implies
    \overbrace{\begin{bmatrix}
        P^{(\sigma)}  & 0\\
        C_j^{(q)}  & 0\\
        0& C^{(q')}_k
    \end{bmatrix} \begin{bmatrix}
        x\\x'
    \end{bmatrix} \leq \begin{bmatrix}
         r^{(\sigma)}\\
         d_j^{(q)} \\
         d^{(q')}_k
    \end{bmatrix}}^{\mathclap{\llangle x \rrangle = \sigma \land V(x,q) \neq \infty \land V(x',q') \neq \infty}}
    \allowdisplaybreaks\\\implies \underbrace{\begin{bmatrix}
       w^{(q)}_j& - w^{(q')}_k
    \end{bmatrix} \begin{bmatrix}
        x\\x'
    \end{bmatrix} \not\leq u^{(q')}_k - u^{(q)}_j + {\bf 1}_F(q, \sigma, q') - 1
    }_{V(x,q) - V(x',q') > {\bf 1}_F(q, \sigma, q') - 1}
\end{multline}
Specifically, the term ${\bf 1}_F(q,\sigma,q') - 1$ evaluates to $-1$ when $(q,\sigma,q') \in \delta \setminus F$, encoding the non-increase requirement of the general inductive step, and to $0$ when $(q,\sigma,q') \in F$, encoding the strict decrease requirement on fair transitions.

Overall, this construction yields $l \cdot |Q_0|$ obligations of the form of \cref{eq:cond-matrix-init}, $n \cdot m \cdot l \cdot |\delta|$ obligations of the form of \cref{eq:cond-matrix-finite}, and $n \cdot m^2 \cdot |\delta|$ obligations of the form of \cref{eq:cond-matrix-step}. For notational convenience, we collapse $[x~ x']$ into a single vector $y$ and observe, assuming appropriate zero padding where necessary, that all obligations can be expressed in the following uniform form:
\begin{align}
    \forall y \,&\colon A_{\sf s}y \leq b_{\sf s} \implies
    C_{\sf p}y \leq d_{\sf p} \implies E_{\sf p} y \not\leq f_{\sf p} \label{eq:obligation}\\
    \Leftrightarrow~
    \lnot \exists y \,&\colon \underbrace{A_{\sf s}y \leq b_{\sf s}}_{\text{secret}} \andimplies \underbrace{C_{\sf p}y \leq d_{\sf p} \andimplies E_{\sf p} y \leq f_{\sf p}}_{\text{public}}\label{eq:unsat}
\end{align}
Equivalently, each derived obligation can be viewed as an unsatisfiability query of the form of \cref{eq:unsat}, consisting of a public and a secret component. Our goal is therefore to convince the verifier that each such query is unsatisfiable, without revealing the secret component.

Each obligation of the form of \cref{eq:unsat} asserts the infeasibility of a system of linear inequalities---a universal statement over all $y$. Rather than proving this directly, we pass to its dual formulation, whose satisfiability certifies the infeasibility of the primal system:
\begin{multline}\label{eq:dual}
    \exists \lambda_{\sf s} \geq 0,\mu_{\sf s} \geq 0:  A_{\sf s}^{\sf T}\lambda_{\sf s} + \overbrace{[C_{\sf p}^{\sf T}~ E^{\sf T}_{\sf p}]}^{G_{\sf p}^{\sf T}}\mu_{\sf s} = 0 \\
    \land b_{\sf s}^{\sf T}\lambda_{\sf s} + \underbrace{[d_{\sf p}^{\sf T} ~ f_{\sf p}^{\sf T}]}_{h_{\sf p}^{\sf T}}\mu_{\sf s} < 0
\end{multline}
By Farkas' lemma, the unsatisfiability of the primal system in \cref{eq:unsat} over the rationals is equivalent to the existence of a rational solution for $\lambda_{\sf s}$ and $\mu_{\sf s}$~\cite{farkas1902theorie}.

The remainder of this section develops our zero-knowledge proof scheme in two steps. We first introduce specialised sub-protocols for the classes of operations that constitute each obligation (\cref{sec:primitives}). We then fold the many obligations of a verification task into a single instance, to which these primitives are applied only once (\cref{sec:folding}).

\subsection{Primitives}\label{sec:primitives}

\begin{figure*}[h]
\centering
    \def\constrwidth{\textwidth}
\begin{minipage}{\constrwidth}
\centering\scriptsize
\setlength{\tabcolsep}{0pt}
\begin{tabular}{|p{0.50\constrwidth} | p{0.50\constrwidth}}
    \hline
    \begin{minipage}[t]{0.50\constrwidth}
    {\phantom{a}\hfill {\bf zkeq.}\textbf{Prove}\hfill\phantom{a}}\\[-4pt]
\noindent\makebox[\linewidth]{\hrulefill}
{\bf Input:} $(\mathsf{sp}, \{(c^{(j)}, r^{(j)}, A^{(j)})\}_{j=1}^{k},(x, c_x, r_x))$\\[-4pt]
\noindent\makebox[\linewidth]{\hrulefill}\\
{Extract bases $\{\hat{g}_i\}_{i=1}^l,  \{\hat{g}_s\}_{s=1}^m, \{\hat{h}^{(j)}\}_{j=1}^k, \ \hat{h}_x$ from $\mathsf{sp}$}\\
$\forall j \in [1,k], \forall i \in [1, l]: \hat{g}^{(j)}_i \gets \prod_{s=1}^{m} (\hat{g}_s)^{A^{(j)}_{si}}$\\
$\forall i \in [1, l],\, r_i \rchoose \mathbb{Z}_p;
\quad \forall j \in [1,k],\, \sigma^{(j)} \rchoose \mathbb{Z}_p; \  \sigma_x \rchoose \mathbb{Z}_p$ \\
$t_x \gets \prod_{i=1}^l \hat{g}_i^{r_i} \cdot \hat{h}_x^{\sigma_x}$\\
$\forall j \in [1,k]: t^{(j)} \gets \prod_{i=1}^l (\hat{g}^{(j)}_i)^{r_i} \cdot (\hat{h}^{(j)})^{\sigma^{(j)}}$\\
$e \gets H(c_x \| c^{(1)} \| \cdots \| c^{(k)} \| t_x \| t^{(1)} \| \cdots \| t^{(k)})$\\
$\forall i \in [1, l]: z_i \gets r_i + e \cdot x_i$\\
$w_x \gets \sigma_x + e \cdot r_x$\\
$\forall j \in [1,k]: w^{(j)} \gets \sigma^{(j)} + e \cdot r^{(j)}$\\
{\bf return} $\pi = (t_x, \{t^{(j)}\}_{j=1}^{k}, \{z_i\}_{i=1}^l, w_x, \{w^{(j)}\}_{j=1}^{k})$\\
    
    \noindent\makebox[\linewidth]{\hrulefill}\\[-2pt]
    {\phantom{a}\hfill {\bf zkrp.}\textbf{Prove}\hfill\phantom{a}}\\[-4pt]
    \noindent\makebox[\linewidth]{\hrulefill}
    {\bf Input:} $(\mathsf{sp}, A, \hat{c}_A=\prod_{i,j=1}^{m,n}\hat{g}_{ij}^{a_{ij}}\hat{h}^{r_A}, r_A, M)$\\[-4pt]
    \noindent\makebox[\linewidth]{\hrulefill}\\
    {Treat $A \in \mathbb{Z}_p^{m \times n}$ as vector $v = (a_{11}, a_{12}, \ldots, a_{mn}) \in \mathbb{Z}_p^l$}\\
    {Extract $g, h, \{g_i\}_{i=1}^{l-1},  \{\hat{g}_{ij}\}_{i,j=1}^{m, n}$, and $g^{\alpha} = g_2$ from $\mathsf{sp}$} \\
    {\tiny (1. Compute Pedersen commitments on  $v$)}\\
    $\forall i \in [1, l], r_i \rchoose \mathbb{Z}_p$, 
    $c_i \gets g^{v_i} h^{r_i}$\\
    {\tiny (2. Prove algebraic linkage between $\{c_i\}_{i=1}^l$ and $\hat{c}_A$)} \\
    Define $f(X)=\sum_{i=1}^{l} v_i X^{i-1}$\\
    $z \gets H(\hat{c}_A\|c_1\|...\|c_{l})$\\
    $r_z=\sum_{i=1}^{l} r_i z^{i-1}$ \\
    $c=\prod_{i=1}^{l} c_i^{z^{i-1}} = g^{f(z)} h^{r_z}$,  \\
    Define $q(X)=\frac{f(X)-f(z)}{X-z}$ \hfill ($q(X)=\sum_{i=1}^{l-1}q_iX^{i-1}$  )\\
    $\mu \overset{R}{\leftarrow} \mathbb{Z}_p$, 
    $\pi_{\sf eq}=\prod_{i=1}^{l-1}g_i^{q_i}h^{\mu}$, 
    $\theta = g^{r_A-r_z+\mu z}(g^{\alpha})^{-\mu}$\\
    {\tiny (3. Prove that each component $v_i$ lies within range $[0, M]$)}\\
    $\pi_{\mathsf{Bullet}} \leftarrow \mathsf{Bullet.Prove}(\mathsf{sp}, \{v_i, c_i, r_i\}_{i=1}^l, M)$\\
    {\bf return} $\pi = (\{c_i\}_{i=1}^l, \pi_{\mathsf{eq}},\theta,\pi_{\mathsf{Bullet}})$\\
    \end{minipage}&
    \parbox[t]{\linewidth}{
    {\phantom{a}\hfill {\bf zkeq.}\textbf{Verify}\hfill\phantom{a}}\\[-4pt]
    \noindent\makebox[\linewidth]{\hrulefill}
    \begin{minipage}[t]{0.50\constrwidth}
    {\bf Input:} $(\mathsf{sp}, \{(c^{(j)}, A^{(j)})\}_{j=1}^{k}, c_x, \pi)$\\[-4pt]
    \noindent\makebox[\linewidth]{\hrulefill}\\
    {Parse $(t_x, \{t^{(j)}\}_{j=1}^{k}, \{z_i\}_{i=1}^l, w_x, \{w^{(j)}\}_{j=1}^{k})$ from $\pi$}\\
    {Extract bases $\{\hat{g}_i\}_{i=1}^l, \{\hat{g}_s\}_{s=1}^m, \{\hat{h}^{(j)}\}_{j=1}^k, \hat{h}_x$ from $\mathsf{sp}$}\\
    $\forall j \in [1,k], \forall i \in [1, l]: \hat{g}^{(j)}_i \gets \prod_{s=1}^{m} (\hat{g}_s)^{A^{(j)}_{si}}$\\
    $e \gets H(c_x \| c^{(1)} \| \cdots \| c^{(k)} \| t_x \| t^{(1)} \| \cdots \| t^{(k)})$\\
    $\forall j \in [1,k]: \prod_{i=1}^l (\hat{g}^{(j)}_i)^{z_i} \cdot (\hat{h}^{(j)})^{w^{(j)}} \overset{?}{=} t^{(j)} \cdot (c^{(j)})^e$\\
    $\prod_{i=1}^l \hat{g}_i^{z_i} \cdot \hat{h}_x^{w_x}
\;\overset{?}{=}\;
t_x \cdot c_x^e$\\
    {\bf accept}\\[-4pt]
    
    \noindent\makebox[\linewidth]{\hrulefill}\\[-2pt]
    {\phantom{a}\hfill {\bf zkrp.}\textbf{Setup}\hfill\phantom{a}}\\[-4pt]
    \noindent\makebox[\linewidth]{\hrulefill}
    {\bf Input:} $(\lambda, ( m,  n))$ with $l= m \cdot  n$\\[-4pt]
    \noindent\makebox[\linewidth]{\hrulefill}\\
    $g \rchoose \mathbb{G}_1,\ g' \rchoose \mathbb{G}_2, \alpha, \beta \rchoose \mathbb{Z}_p$\\
    $\forall i \in [1,  l-1]:\ g_i \gets g^{\alpha^{i-1}}$\\
    
    $h = g^\beta,\ h' = g'^{\beta},\ \hat{h} = e(h, g')$\\
    $\forall i \in [1,  m], j \in [1,  n]:\ \hat{g}_{ij} = e(g^{\alpha^{j-1}}, g'^{\alpha^{n(i-1)}})$\\
    ${\sf Bullet.sp}\leftarrow {\sf Bullet.setup}(\lambda)$\\
    {\bf return} $\mathsf{sp}=( m,  n, g, g', g'^\alpha, h, h', \hat{h}, \{g_i\}_{i=1}^{l-1},
    \{\hat{g}_{ij}\}_{i,j=1}^{ m, n}, \mathsf{Bullet}.\mathsf{sp})$\\
    
    \noindent\makebox[\linewidth]{\hrulefill}\\[-2pt]
    {\phantom{a}\hfill {\bf zkrp.}\textbf{Verify}\hfill\phantom{a}}\\[-4pt]
    \noindent\makebox[\linewidth]{\hrulefill}
    {\bf Input:} $(\mathsf{sp}, \hat{c}_A, M, \pi)$\\[-4pt]
    \noindent\makebox[\linewidth]{\hrulefill}\\
    {Parse $\pi = (\{c_i\}_{i=1}^l,  \pi_{\mathsf{eq}}, \theta, \pi_{\mathsf{Bullet}})$}\\
    {Extract bases $g', g'^\alpha,  h'$ from $\mathsf{sp}$}\\
    $z \gets H(\hat{c}_A \| c_1 \| ... \|c_l)$\\
    $c = \prod_{i=1}^{l} c_i^{z^{i-1}}$\\
    $\hat{c}_A \cdot e(c, g')^{-1} \overset{?}{=} e(\pi_{\mathsf{eq}}, g'^{\alpha}g'^{-z}) \cdot e(\theta, h')$\\
    $\mathsf{Bullet.Verify}(\mathsf{sp}, \{c_i\}_{i=1}^l, M, \pi_{\mathsf{Bullet}}) \overset{?}{=} 1$\\
    
    {\bf accept}
    \end{minipage}
    }%
    \hspace{-0.4pt}\vrule\\
    \hline
\end{tabular}
\end{minipage}

\caption{Primitives of our Symbolic ZKMC Scheme}
\label{fig:ZKMC-S2-routines}
\end{figure*}

While Farkas' lemma guarantees the existence of rational witnesses, our cryptographic routines operate over a prime-order field $\mathbb{Z}_p$.
We use the standard bounded encoding of signed integers~\cite{couteau2021efficient,hader2024smt}: for a bound $M$ with $M^2 \ll p$, the range $[0, M]$ represents the non-negative integers and $[p - M, p - 1]$ represents the negatives $[-M, -1]$, so that field arithmetic preserves integer semantics without wraparound on values whose magnitudes stay below $\sqrt{p}$.
For our benchmarks, a 255-bit prime with $M = 2^{32}$ keeps the relevant products $A_{\sf s}^{\sf T} \lambda_{\sf s}$ and $b_{\sf s}^{\sf T} \lambda_{\sf s}$ below $\hat n' \cdot M^2 \approx 2^{74} \ll p$, where $\hat n'$ is the number of rows of $A_{\sf s}$.
Our certification scheme therefore computes two secret integer Farkas witnesses $\lambda_{\sf s}, \mu_{\sf s} \in [0, M]$.
Their existence suffices to establish the corresponding obligation in \cref{eq:obligation} but is not necessary: the dual system may admit only rational or out-of-bound solutions, which affects the completeness of our approach but not its soundness.
Accordingly, our scheme requires the prover to demonstrate in zero knowledge that the secret certificates satisfy:
\tightmath
\begin{align}\label{eq:farkas-zk-proof}
    &\lambda_{\sf s}, \mu_{\sf s} \in [0, M] \land
    \overbrace{-b_{\sf s}^{\sf T} \lambda_{\sf s} - h_{\sf p}^{\sf T} \mu_{\sf s} - 1 \in [0, M] \land
    A_{\sf s}^{\sf T} \lambda_{\sf s}  = - G_{\sf p}^{\sf T} \mu_{\sf s}}^{\text{matrix multiplication}}\notag\\[-3ex]
    &\underbrace{\phantom{\smash{\lambda_{\sf s},\mu_{\sf s} \in [0, M] \land
    -b_{\sf s}^{\sf T} \lambda_{\sf s} - h_{\sf p}^{\sf T} \mu_{\sf s} - 1 \in [0, M]}}}_{\text{range proof}}
    \phantom{\smash{\land}}
    \underbrace{\phantom{\smash{A_{\sf s}^{\sf T} \lambda_{\sf s}  = - G_{\sf p}^{\sf T} \mu_{\sf s}}}}_{\text{equality proof}}
\end{align}\normalmath

Each obligation of the form of \cref{eq:farkas-zk-proof} comprises three classes of operations: (i) matrix--vector multiplications and additions, (ii) range membership proofs, and (iii) equality proofs. This decomposition suggests a modular approach that favours prover efficiency: we design a specialised zero-knowledge sub-protocol for each class, prove each independently and then compose them into a single zero-knowledge proof.

Since we operate over matrices and vectors, we need a commitment scheme that exposes their row and column structure to the sub-protocols.
We meet this requirement with the \emph{two-tier} Pedersen vector commitment of~\cite{CongYY24}, which lifts the standard Pedersen commitment~\cite{DBLP:conf/crypto/Pedersen91} to matrices via a bilinear pairing.
The standard scheme commits to a scalar $v \in \mathbb{Z}_p$ under randomness $r \in \mathbb{Z}_p$ as $\mathsf{Com}(v; r) = g^v h^r$ for public generators $g, h \in \mathbb{G}_1$; it is perfectly hiding, computationally binding under the discrete-logarithm assumption, and additively homomorphic: $\mathsf{Com}(v; r) \cdot \mathsf{Com}(v'; r') = \mathsf{Com}(v + v'; r + r')$.
The vector extension uses independent generators $g_1, \dots, g_n$ to commit to $\mathbf{v} \in \mathbb{Z}_p^n$ as $\mathsf{Com}(\mathbf{v}; r) = h^r \prod_{i=1}^n g_i^{v_i}$, preserving all three properties.

The matrix extension, in turn, works over a bilinear group $(\mathbb{G}_1, \mathbb{G}_2, \mathbb{G}_T)$ of prime order $p$ with pairing $e\colon \mathbb{G}_1 \times \mathbb{G}_2 \to \mathbb{G}_T$~\cite{DBLP:conf/asiacrypt/BonehLS01,DBLP:journals/dam/GalbraithPS08}.
Bilinearity, $e(g^a, g'^b) = e(g, g')^{ab}$, lets us aggregate column-level commitments in $\mathbb{G}_1$ into a single matrix-level commitment in $\mathbb{G}_T$ that still binds to each column.
Concretely, each column $\mathbf{a}_j$ of $A \in \mathbb{Z}_p^{m \times n}$ is committed as a vector commitment $c_j \in \mathbb{G}_1$, and the columns are then aggregated into a matrix commitment $\hat c_A = \prod_{j=1}^n e(c_j, g'_j) \in \mathbb{G}_T$ for public generators $g'_j \in \mathbb{G}_2$.
We can thus open at the column level (via $c_j$) or at the matrix level (via $\hat c_A$) without re-committing.

Finally, to avoid the round-trip cost of interaction, we apply the Fiat--Shamir transform~\cite{FiatS86} in the random oracle model: each sub-protocol derives its challenges by hashing the prover's first messages with a hash function $H$, which the verifier can replay locally.

\paragraph{Matrix Multiplication and Addition}
Our protocol requires proofs of consistency for committed matrix products. To realise them, we adopt the zero-knowledge matrix multiplication protocol of zkMatrix~\cite{CongYY24} and only state its interface. It provides the routine $\mathsf{zkmm.Prove}(\mathsf{sp}, A, B, c_A, c_B, r_A, r_B; r_{AB}) \to (\pi_{AB}, c_{AB})$, which outputs a commitment $c_{AB}$ to the product $AB$ together with a proof $\pi_{AB}$ that it is consistent with the committed operands, verified by $\mathsf{zkmm.Verify}(\mathsf{sp}, c_A, c_B, c_{AB}, \pi_{AB}) \to \{0,1\}$. The scheme implements the following relation:
\begin{equation*}
    \mathcal{R}_{\mathsf{zkmm}} = \pok \begin{Bmatrix}
    (A \in \mathbb{Z}_p^{m \times l}, B \in \mathbb{Z}_p^{l \times n}, r_A, r_B, r_{AB}) ; \\
    c_A, c_B, c_{AB} : \\
    c_A = \mathsf{Com}(\mathsf{sp}, A; r_A) \land \\
    c_B = \mathsf{Com}(\mathsf{sp}, B; r_B) \\
    \land\ c_{AB} = \mathsf{Com}(\mathsf{sp}, AB; r_{AB})
    \end{Bmatrix}
\end{equation*}
Additions, by contrast, require no dedicated protocol: since the commitment scheme is additively homomorphic, proving and verifying a commitment to an addition can be carried out solely using the commitment algorithms.

\paragraph{Equality Proof}
Our protocol requires two equality proof operations of the form $y_{\sec} = A_{\pub} x_{\sec}$, where $A_{\pub}$ is public and $x_{\sec}, y_{\sec}$ are secret: the first corresponds to the proof obligation $A_{\sec}^{\top} \lambda_{\sec} = -G_{\pub}^{\top} \mu_{\sec}$ appearing in \cref{eq:farkas-zk-proof}, and the second arises from the commitment $c_{\gamma}$ to $-h_{\pub}^{\top} \mu_{\sec}$ required by our protocol construction (see \cref{fig:ZKMC-S2-fold}). We batch both operations using the routine
$
\mathsf{zkeq.Prove}\bigl(\mathsf{sp},\allowbreak \{(c^{(j)}, r^{(j)}, A^{(j)})\}_{j=1}^{k},\allowbreak (x, c_x, r_x)\bigr) \to \pi,
$
which outputs a proof that, for each $j$, the commitment $c^{(j)}$ with randomness $r^{(j)}$ opens to $A^{(j)} x$, where $x$ itself is committed as $c_x$ using randomness $r_x$. The verification algorithm
$
\mathsf{zkeq.Verify}\bigl(\mathsf{sp}, \allowbreak \{(c^{(j)}, A^{(j)})\}_{j=1}^{k}, \allowbreak c_x, \allowbreak \pi\bigr)
$
accepts if and only if all commitments are consistent. The scheme implements the following relation:
\begin{equation*}
\mathcal{R}_{\mathsf{zkeq}} = \pok \begin{Bmatrix}
(x, r_x, \{r^{(j)}\}_{j=1}^{k}) ; \\
\{A^{(j)}\}_{j=1}^{k}, c_x, \{c^{(j)}\}_{j=1}^{k} : \\
\forall j\mathpunct.\! c^{(j)} = \mathsf{Com}(\mathsf{sp}, A^{(j)} x; r^{(j)}) \\\land c_x = \mathsf{Com}(\mathsf{sp}, x; r_x)
\end{Bmatrix}
\end{equation*}
Conceptually, $\mathsf{zkeq}$ encodes the public matrices $A^{(j)}$ into the commitment bases and reduces the statement to a proof of discrete-logarithm equality~\cite{ChaumP92}.

\paragraph{Range Proof}
Our obligation requires positivity checks over the range $[0, M]$, as illustrated in \cref{eq:farkas-zk-proof}, and bound checks $A_{\sec}, b_{\sec} \in [-M, M]$ on signed coefficients. We reduce the signed checks to non-negative ones via the offset $A_{\sec} + M \in [0, 2M]$ and $b_{\sec} + M \in [0, 2M]$ and prove that every entry of a committed matrix lies in $[0, M]$, captured by the following relation:
\begin{equation*}
    \mathcal{R}_{\mathsf{zkrp}} = \pok \begin{Bmatrix}
    (A \in \mathbb{Z}_p^{m \times n}, r_A) ; c_A, M : \\
    c_A = \mathsf{Com}(\mathsf{sp}, A; r_A) \land \forall i, j\colon a_{ij} \in [0, M]
    \end{Bmatrix}
\end{equation*}
To realise this relation, we provide the routine \textsf{zkrp} (\Cref{fig:ZKMC-S2-routines}), with prover $\mathsf{zkrp.Prove}(\mathsf{sp},\allowbreak A,\allowbreak c_A,\allowbreak r_A,\allowbreak M) \to \pi$ and verifier $\mathsf{zkrp.Verify}(\mathsf{sp}, c_A, M, \pi) \to \{0, 1\}$.
We construct \textsf{zkrp} by flattening $A$ and committing to each entry under an individual Pedersen commitment $c_i$, then discharging two sub-tasks: per-element range checks on the $c_i$, and a consistency check that ties the $\{c_i\}$ back to the matrix commitment $\hat c_A$.
The per-element non-negative checks are discharged by Bulletproofs~\cite{bunz2018bulletproofs}, a logarithmic-size zero-knowledge proof for proving range membership of values committed under Pedersen commitments: a prover $\mathsf{Bullet.Prove}(\mathsf{sp}, \{v_i, r_i, c_i\}_{i=1}^{k}, M) \to \pi$ and a verifier $\mathsf{Bullet.Verify}(\mathsf{sp}, \{c_i\}_{i=1}^{k}, M, \pi) \to \{0,1\}$ establish that each $c_i = g^{v_i} h^{r_i}$ commits to a value $v_i \in [0, M]$.
It remains to certify that the per-entry commitments $\{c_i\}$ and the matrix commitment $\hat c_A$ encode the same data. We adapt the KZG opening technique~\cite{KateZG10}: we encode the entries as the coefficients of a polynomial $f$ whose exponents follow the matrix layout, so that an evaluation of $f$ at a verifier-chosen challenge $z$ uniquely determines the underlying matrix. The prover then produces a randomised KZG-style opening proof for $f(z)$ together with a blinding-correction term, and a single pairing equation certifies that $\hat c_A$ and $\{c_i\}$ are consistent with this opening.
The resulting protocol is complete, knowledge-sound in the algebraic group model with a random oracle, and zero-knowledge in the random oracle model; we sketch the argument in \Cref{sec:zkrpsecurity}.

\subsection{Folding}\label{sec:folding}

The primitives of \cref{sec:primitives} produce one obligation at a time, and a
verification task can produce thousands of them
(\cref{tab:experimental_results}).
Every obligation, however, has the same shape and differs only in its data.
We exploit this uniformity and \emph{fold} all obligations of a task into a single
instance, so that the prover runs the matrix multiplication sub-protocols once
instead of $N$ times.
We index the obligations by $i \in [N]$ and abbreviate
$A_i = A_{\sec,i}^\T$, $b_i = -b_{\sec,i}^\T$, $G_i = -G_{\pub,i}^\T$ and
$h_i = -h_{\pub,i}^\T$, so that the multiplicative part of
\cref{eq:farkas-zk-proof} becomes two secret--secret and two public--secret
families of matrix--vector products:
\begin{equation}\label{eq:fold-families}
    \underbrace{A_i \lambda_i = \alpha_i,\ \ b_i \lambda_i = \beta_i}_{\text{secret--secret}}
    \qquad
    \underbrace{G_i \mu_i = \alpha_i,\ \ h_i \mu_i = \gamma_i}_{\text{public--secret}}
\end{equation}
Both groups share $\alpha_i$, which is what ties
$A_{\sec,i}^\T \lambda_{\sec,i} = -G_{\pub,i}^\T \mu_{\sec,i}$ together.
The remaining conditions $\lambda_i, \mu_i, \delta_i \in [0,M]$ and
$A_i, b_i \in [-M,M]$ constrain ranges, where $\delta_i = \beta_i + \gamma_i - 1$
is the second range term of \cref{eq:farkas-zk-proof}.

\begin{figure*}[h]
\centering
\def\constrwidth{\textwidth}
\begin{minipage}{\constrwidth}
\centering\scriptsize
\setlength{\tabcolsep}{0pt}
\begin{tabular}{|p{\constrwidth}|}
    \hline
    \textbf{Prover Input:} Secret matrices $A_i = A_{\sec,i}^\T$ and vectors $b_i = -b_{\sec,i}^\T$, randomness $r_{A_i}, r_{b_i}$, for every obligation $i \in [N]$\\
    \textbf{Common Input:} Security parameter $1^\lambda$, public matrices $G_i = -G_{\pub,i}^\T$ and vectors $h_i = -h_{\pub,i}^\T$, upper bound $M$, setup $\mathsf{sp}$ as in \cref{fig:ZKMC-S2-routines},
    exponent schedule $C = \{c_1, \dots, c_N\}$ with $2C \cap K = \varnothing$, where $K = \{c_i + c_j \mid i \neq j\}$,
    offset commitments $c_\theta, c'_\theta$ to the all-$M$ matrix/vector (randomness $0$), shifting $A_i, b_i$ into $[0, 2M]$
    \\
    \hline
\end{tabular}
\begin{tabular}
{|p{0.55\constrwidth} | p{0.45\constrwidth}}
    \centering\textbf{Prover} & \hfill\centering\textbf{Verifier} \hfill\vrule\tabularnewline
    \hline
    \begin{minipage}[t]{0.55\constrwidth}
    Compute Farkas witnesses $\lambda_i, \mu_i$ for all $i \in [N]$\\
    {\bf sample} $r_{\lambda_i}, r_{\mu_i}, r_{\alpha_i}, r_{\beta_i}, r_{\gamma_i}$\\
    $\forall i$: $\alpha_i \gets A_i \lambda_i$,~ $\beta_i \gets b_i \lambda_i$,~ $\gamma_i \gets h_i \mu_i$\\
    $\forall i$: commit to $A_i, b_i, \lambda_i, \mu_i, \alpha_i, \beta_i, \gamma_i$\hfill{\tiny(two-tier Pedersen, \cref{sec:primitives})}\\
    {\bf send} ${\cal C}_0 = \{(c_{A_i}, c_{b_i}, c_{\lambda_i}, c_{\mu_i}, c_{\alpha_i}, c_{\beta_i}, c_{\gamma_i})\}_{i \in [N]}$\\[-2pt]
    \noindent\makebox[\linewidth]{\dotfill}\\
    $\forall k \in K$: $C^A_k \gets \sum_{c_i + c_j = k,\, i \neq j} A_i \lambda_j$\hfill{\tiny(aggregated cross terms)}\\
    \phantom{$\forall k \in K$: }~and $C^b_k, C^G_k, C^h_k$ likewise from $b_i \lambda_j$, $G_i \mu_j$, $h_i \mu_j$\\
    {\bf sample} $\omega^A_k, \omega^b_k, \omega^G_k, \omega^h_k$;\quad $c_{C^X_k} \gets \mathsf{Commit}(\mathsf{sp}, C^X_k; \omega^X_k)$\\
    {\bf send} $\{(c_{C^A_k}, c_{C^b_k}, c_{C^G_k}, c_{C^h_k})\}_{k \in K}$\\[1pt]
    $r \gets H({\cal C}_0 \| \{c_{C^A_k}, c_{C^b_k}, c_{C^G_k}, c_{C^h_k}\}_{k \in K})$\\
    $A^* \gets \sum_i r^{c_i} A_i$,\quad $b^* \gets \sum_i r^{c_i} b_i$\hfill{\tiny(fold the witnesses)}\\
    $\lambda^* \gets \sum_i r^{c_i} \lambda_i$,\quad $\mu^* \gets \sum_i r^{c_i} \mu_i$\\
    $r_{A^*} \gets \sum_i r^{c_i} r_{A_i}$, and $r_{b^*}, r_{\lambda^*}, r_{\mu^*}$ likewise\\[1pt]
    $\alpha^*_A \gets \sum_i r^{2c_i} \alpha_i + \sum_{k \in K} r^k C^A_k$\hfill{\tiny($= A^* \lambda^*$)}\\
    $\alpha^*_G, \beta^*, \gamma^*$ likewise from $C^G_k, C^b_k, C^h_k$\hfill{\tiny($= G^*\mu^*,\, b^*\lambda^*,\, h^*\mu^*$)}\\
    $r_{\alpha^*_A} \gets \sum_i r^{2c_i} r_{\alpha_i} + \sum_{k \in K} r^k \omega^A_k$, and $r_{\alpha^*_G}, r_{\beta^*}, r_{\gamma^*}$ likewise\\[2pt]
    $(\pi_\alpha, \cdot) \gets \mathsf{zkmm.Prove}(\mathsf{sp}, A^*, \lambda^*, c_{A^*}, c_{\lambda^*}, r_{A^*}, r_{\lambda^*}; r_{\alpha^*_A})$\\
    $(\pi_\beta, \cdot) \gets \mathsf{zkmm.Prove}(\mathsf{sp}, b^*, \lambda^*, c_{b^*}, c_{\lambda^*}, r_{b^*}, r_{\lambda^*}; r_{\beta^*})$\\
    $\pi_{\eta} \gets \mathsf{zkeq.Prove}(\mathsf{sp}, \{(c_{\alpha^*_G}, r_{\alpha^*_G}, G^*), (c_{\gamma^*}, r_{\gamma^*}, h^*)\}, (\mu^*, c_{\mu^*}, r_{\mu^*}))$\\[1pt]
    $\pi_{\sf rp} \gets \mathsf{zkrp.Prove}$ on $\{c_{\lambda_i}, c_{\mu_i}, c_{\delta_i}, c_{A_i} c_\theta, c_{b_i} c'_\theta\}_{i \in [N]}$\hfill{\tiny(batched)}\\
    {\bf send} $\pi = (\pi_\alpha, \pi_\beta, \pi_\eta, \pi_{\sf rp})$\\[-3pt]
    \noindent\makebox[\linewidth]{\hrulefill}
    \end{minipage}&
    \parbox[t]{\linewidth}{
    \begin{minipage}[t]{0.45\constrwidth}
    {\bf receive} ${\cal C}_0$\\[-2pt]
    \noindent\makebox[\linewidth]{\dotfill}
    {\bf receive} $\{(c_{C^A_k}, c_{C^b_k}, c_{C^G_k}, c_{C^h_k})\}_{k \in K}$ and $\pi$\\
    $r \gets H({\cal C}_0 \| \{c_{C^A_k}, c_{C^b_k}, c_{C^G_k}, c_{C^h_k}\}_{k \in K})$\\
    $G^* \gets \sum_i r^{c_i} G_i$,\quad $h^* \gets \sum_i r^{c_i} h_i$\\
    $c_{A^*} \gets \prod_i c_{A_i}^{r^{c_i}}$, and $c_{b^*}, c_{\lambda^*}, c_{\mu^*}$ likewise\\
    $c_{\alpha^*_A} \gets \prod_i c_{\alpha_i}^{r^{2c_i}} \cdot \prod_{k \in K} c_{C^A_k}^{r^k}$\\
    $c_{\alpha^*_G}, c_{\beta^*}, c_{\gamma^*}$ likewise from $c_{C^G_k}, c_{C^b_k}, c_{C^h_k}$\\[1pt]
    $\mathsf{zkmm.Verify}(\mathsf{sp}, c_{A^*}, c_{\lambda^*}, c_{\alpha^*_A}, \pi_\alpha) \overset{?}{=} 1$\\
    $\mathsf{zkmm.Verify}(\mathsf{sp}, c_{b^*}, c_{\lambda^*}, c_{\beta^*}, \pi_\beta) \overset{?}{=} 1$\\
    $\mathsf{zkeq.Verify}(\mathsf{sp}, \{(c_{\alpha^*_G}, G^*), (c_{\gamma^*}, h^*)\}, c_{\mu^*}, \pi_\eta) \overset{?}{=} 1$\\[1pt]
    $\forall i$: $c_{\delta_i} \gets c_{\beta_i} \cdot c_{\gamma_i} \cdot \mathsf{Commit}(\mathsf{sp}, -1; 0)$\\
    $\mathsf{zkrp.Verify}$ on $\{c_{\lambda_i}, c_{\mu_i}, c_{\delta_i}, c_{A_i} c_\theta, c_{b_i} c'_\theta\}_{i \in [N]} \overset{?}{=} 1$\\
    \textbf{accept}\\[-4pt]
    \noindent\makebox[\linewidth]{\hrulefill}
    \end{minipage}
    }%
    \hspace{-0.4pt}\vrule
\end{tabular}
\end{minipage}

\caption{Construction of our Folding ZKMC Scheme}
\label{fig:ZKMC-S2-fold}
\end{figure*}

Our folding protocol realises the proof of knowledge below, which
establishes all $N$ proof obligations at once:
\tightmath
\begin{equation*}
    {\cal R}_{\sf f} = \pok \begin{Bmatrix}
    (\{A_i, b_i, \lambda_i, \mu_i, r_{A_i}, r_{b_i}, r_{\lambda_i}, r_{\mu_i}\}_{i \in [N]}) ; \\
    (\{G_i, h_i, c_{A_i}, c_{b_i}, c_{\lambda_i}, c_{\mu_i}\}_{i \in [N]}, M) : \\
    \forall i \in [N] \colon
    A_i \lambda_i = G_i \mu_i \land \\
    \lambda_i, \mu_i \in [0, M] \land
    b_i \lambda_i + h_i \mu_i - 1 \in [0, M] \land \\
    A_i \in [-M, M] \land b_i \in [-M, M] \land \\
    \mathsf{Com}({\sf sp}, A_i; r_{A_i}) = c_{A_i} \land \\
    \mathsf{Com}({\sf sp}, b_i; r_{b_i}) = c_{b_i} \land \\
    \mathsf{Com}({\sf sp}, \lambda_i; r_{\lambda_i}) = c_{\lambda_i} \land \\
    \mathsf{Com}({\sf sp}, \mu_i; r_{\mu_i}) = c_{\mu_i}
    \end{Bmatrix}
\end{equation*}\normalmath
Each family of \cref{eq:fold-families} comprises $N$ instances of a single
relation, and $\mathsf{zkmm}$ proves one instance at a time.
We thus combine the $N$ instances into a single pair
of operands, summing individual instances under random scalars.
The natural candidate weights instance $i$ by a power $r^{c_i}$ of a challenge $r$
that the prover derives by Fiat--Shamir, as in \cref{sec:primitives}, and sums the
result into $A^* = \sum_i r^{c_i} A_i$ and $\lambda^* = \sum_i r^{c_i} \lambda_i$.
The product of the two is:
\tightmath
\begin{equation}\label{eq:fold-expand}
    A^* \lambda^* =
    \underbrace{\sum_i r^{2 c_i} A_i \lambda_i}_{\text{diagonal terms}}
    + \underbrace{\sum_{i \neq j} r^{c_i + c_j} A_i \lambda_j}_{\text{cross terms}}
\end{equation}\normalmath
The diagonal terms correspond to the instances we set out to prove, each carried by its own
power $r^{2 c_i}$.
The cross terms multiply data from different obligations and correspond to nothing
in \cref{eq:fold-families}, yet the prover cannot discard them, because the
verifier checks the product that it computes from the folded commitments.
Folding schemes~\cite{nova,protogalaxy} resolve this tension by absorbing the cross
terms into the statement: the prover commits to them before it learns the
challenge, and both parties then define the folded right-hand side to include
them.
Since only the sum of the cross terms sharing a power of $r$ enters the check, one
commitment per distinct exponent suffices.
Writing $K = \{c_i + c_j \mid i \neq j\}$ for the set of cross-term exponents, the
prover sends a commitment to
$C^A_k = \sum_{c_i + c_j = k,\, i \neq j} A_i \lambda_j$ for every $k \in K$, and
both parties take
$\alpha^*_A = \sum_i r^{2 c_i} \alpha_i + \sum_{k \in K} r^k C^A_k$
as the folded right-hand side, which by \cref{eq:fold-expand} equals
$A^* \lambda^*$ exactly when every instance holds.
The three remaining families fold in the same way.
Every step combines committed values linearly, so the verifier can recompute the
folded commitments itself from the per-obligation commitments and the $4|K|$
cross-term commitments (via the homomorphic properties), and folds the public $G^*$ and $h^*$ in plaintext.
One invocation of $\mathsf{zkmm}$ per secret--secret family and one of $\mathsf{zkeq}$
for the two public--secret families then replace the $N$ per-obligation invocations;
\cref{fig:ZKMC-S2-fold} shows the resulting construction, which realises ${\cal R}_{\sf f}$.

It remains to choose the exponents $c_i$: we call their public assignment
$C = \{c_1, \dots, c_N\}$ the \emph{exponent schedule}, and its choice decides
whether folding is sound at all.
Committing to a cross term binds the prover to that term but does not constrain
its contents, as nothing requires $C^A_k$ to hold the sum that it should.
\Cref{eq:fold-expand} can be seen as a polynomial identity in the challenge: the
coefficient at $X^{2 c_i}$ is the residual $A_i \lambda_i - \alpha_i$ of obligation
$i$, and the coefficient at $X^k$ with $k \in K$ is the residual of the
corresponding aggregate.
If a diagonal exponent $2 c_i$ would coincide with a cross-term exponent, a
dishonest prover could cancel the residual of obligation $i$ against the aggregate
that it chose freely at the same power, and the identity would hold for
\emph{every} challenge.
An admissible schedule must therefore separate the two sets of exponents:
\begin{equation}\label{eq:schedule}
    2C \cap K = \varnothing,
    \quad\text{i.e.}\quad
    2 c_i \neq c_j + c_k \text{ for all } i \text{ and all } j \neq k
\end{equation}
A violated obligation then leaves a non-zero polynomial of degree at most
$2 \max_i \{c_i\}$, which the Schwartz--Zippel
lemma~\cite{schwartz1980,zippel1979} confines to a $2 \max_i \{c_i\} / p$ fraction of
the challenges.
Collisions \emph{within} $K$ remain harmless, which is what permits the
aggregation above.
Condition~\cref{eq:schedule} states precisely that $C$ contains no three-term
arithmetic progression (i.e., $C$ is a \emph{Salem--Spencer
set}~\cite{erdos1936,salem1942,behrend1946}) and thereby excludes the
obvious schedule $c_i = i - 1$; reading the binary digits of $i-1$ in base $3$, by
contrast, yields an admissible schedule with $|K| < 3^{\lceil \log_2 N \rceil}$.

\begin{table*}[h]
    \centering
    \newcolumntype{C}[1]{>{\centering\arraybackslash}p{#1}}
    \def\colsz{10mm}
    \setlength{\tabcolsep}{3pt}
    \caption{Runtime and problem size comparison (OOT: > \SI{2}{\hour}; ``-'': not run; OOM: out of memory).}
    \begin{tabular}{ |l|l|C{\colsz}|C{\colsz}|C{\colsz}|C{\colsz}|C{\colsz}|C{\colsz}|C{\colsz}|C{\colsz}|C{\colsz}|C{\colsz}| }
    \cline{3-12}
     \multicolumn{2}{c|}{} & \multicolumn{5}{c|}{Explicit-State ZKMC} & \multicolumn{5}{c|}{Symbolic (Folding) ZKMC}\\
     \hline
               &  &  & \multicolumn{4}{c|}{Runtime} & Oblig- & \multicolumn{4}{c|}{Runtime} \\\cline{4-7} \cline{9-12}
     Benchmark & \multicolumn{1}{c|}{$|S|$} & $\sum_\star|E_\star|$ & Enum. & Setup & Prover & Verifier & ations & Farkas & Setup & Prover & Verifier \\
     \hline
    \texttt{exb\_i1a2} & \zkroundspace{32} & \zkroundspace{104} & \zkroundtime{0.14} & \zkroundtime{0.04} & \zkroundtime{0.02} & \zkroundtime{0.02} & 217 & \zkroundtime{2.67} & \zkroundtime{0.63} & \zkroundtime{78.788} & \zkroundtime{9.655}\\
    \texttt{exb\_i2a2} & \zkroundspace{64} & \zkroundspace{400} & \zkroundtime{0.42} & \zkroundtime{0.08} & \zkroundtime{0.04} & \zkroundtime{0.02} & 217 & \zkroundtime{2.66} & \zkroundtime{0.63} & \zkroundtime{78.741} & \zkroundtime{9.649}\\
    \texttt{exb\_i4a2} & \zkroundspace{128} & \zkroundspace{1568}  & \zkroundtime{1.55} & \zkroundtime{0.26} & \zkroundtime{0.12} & \zkroundtime{0.04} & 217 & \zkroundtime{2.67} & \zkroundtime{0.62} & \zkroundtime{78.875} & \zkroundtime{9.653}\\
    \texttt{exb\_i2a3} & \zkroundspace{192} & \zkroundspace{3504} & \zkroundtime{3.63} & \zkroundtime{0.87} & \zkroundtime{0.38} & \zkroundtime{0.06} & 260 & \zkroundtime{3.23} & \zkroundtime{0.62} & \zkroundtime{78.757} & \zkroundtime{9.646}\\
    \texttt{exb\_i8a2} & \zkroundspace{256} & \zkroundspace{6208} & \zkroundtime{6.05} & \zkroundtime{0.87} & \zkroundtime{0.38} & \zkroundtime{0.07} & 217 & \zkroundtime{2.67} &  \zkroundtime{0.62} & \zkroundtime{78.672} & \zkroundtime{9.647}\\
    \texttt{exb\_i4a3} & \zkroundspace{384} & \zkroundspace{13920} & \zkroundtime{14.42} & \zkroundtime{3.1} & \zkroundtime{1.5} & \zkroundtime{0.16} & 260 & \zkroundtime{3.23} & \zkroundtime{0.63} & \zkroundtime{78.807} & \zkroundtime{9.634}\\
    \texttt{exb\_i8a3} & \zkroundspace{768} & \zkroundspace{55488} & \zkroundtime{57.35} & \zkroundtime{11.5} & \zkroundtime{5.7} & \zkroundtime{0.59} & 260 & \zkroundtime{3.23} & \zkroundtime{0.63} & \zkroundtime{78.796} & \zkroundtime{9.634}\\
    \texttt{exb\_i64a3} & \zkroundspace{6144} & \zkroundspace{3540480} & \zkroundtime{3885.652} & \zkroundtime{627.453} & OOM & - & 260 & \zkroundtime{3.24} & \zkroundtime{0.62} & \zkroundtime{78.827} & \zkroundtime{9.643}\\
    \texttt{exb\_i128a3} & \zkroundspace{12288} & - & OOT & - & - & - & 260 & \zkroundtime{3.24} & \zkroundtime{0.62} & \zkroundtime{78.897} & \zkroundtime{9.641}\\
    \texttt{exb\_i64a4} & \zkroundspace{16384} & - & - & - & - & - & 303 & \zkroundtime{3.8} & \zkroundtime{0.62} & \zkroundtime{78.674} & \zkroundtime{9.638}\\
    \texttt{exb\_i128a4} & \zkroundspace{32768} & - & - & - & - & - & 303 & \zkroundtime{3.8} & \zkroundtime{0.62} & \zkroundtime{78.791} & \zkroundtime{9.641}\\
    \hhline{|=|=|=|=|=|=|=|=|=|=|=|=|}
    \texttt{dhcp\_noOFF\_7\_2\_7} & \zkroundspace{1120} & \zkroundspace{439072} & \zkroundtime{163.709} & \zkroundtime{11.5} & \zkroundtime{5.6} & \zkroundtime{4.4} & 1649 & \zkroundtime{25.439} & \zkroundtime{2.038} & \zkroundtime{921.904} & \zkroundtime{92.799}\\
    \texttt{dhcp\_noOFF\_100\_2\_10} & \zkroundspace{11312} & - & OOT & - & - & - & 1649 & \zkroundtime{24.648} & \zkroundtime{2.031} & \zkroundtime{921.874} & \zkroundtime{92.763}\\
    \texttt{dhcp\_noOFF\_1000\_2\_10} & \zkroundspace{112112} & - & - & - & - & - & 1649 & \zkroundtime{25.012} & \zkroundtime{2.036} & \zkroundtime{922.041} & \zkroundtime{92.758}\\
    \hline
    \texttt{dhcp\_7\_2\_7} & \zkroundspace{1280} & \zkroundspace{584224} & \zkroundtime{228.609} & \zkroundtime{21.7} & \zkroundtime{11.9} & \zkroundtime{6.2} & 2139 & \zkroundtime{32.520} & \zkroundtime{3.177} & \zkroundtime{1174.944} & \zkroundtime{117.935}\\
    \texttt{dhcp\_noOFF\_7\_3\_7} & \zkroundspace{3024} & \zkroundspace{3000000} & \zkroundtime{1337.272} & \zkroundtime{82.4} & \zkroundtime{45.6} & \zkroundtime{30.4} & 2252 & \zkroundtime{34.898} & \zkroundtime{3.256} & \zkroundtime{1231.164} & \zkroundtime{123.794}\\
    \hline
    \texttt{dhcp\_7\_3\_7} & \zkroundspace{3456} & \zkroundspace{4070136} & \zkroundtime{1838.354} & \zkroundtime{158.8} & \zkroundtime{88.3} & \zkroundtime{38.462} & 2847 & \zkroundtime{43.293} & \zkroundtime{3.696} & \zkroundtime{1526.737} & \zkroundtime{153.873}\\
    \texttt{dhcp\_100\_3\_10} & \zkroundspace{19392} & - & OOT & - & - & - & 2847 & \zkroundtime{43.489} & \zkroundtime{3.708} & \zkroundtime{1525.890} & \zkroundtime{153.869}\\
    \texttt{dhcp\_1000\_3\_10} & \zkroundspace{192192} & - & - & - & - & - & 2847 & \zkroundtime{43.381} & \zkroundtime{3.687} & \zkroundtime{1526.610} & \zkroundtime{153.879}\\
    \texttt{dhcp\_10000\_3\_10} & \zkroundspace{1920192} & - & - & - & - & - & 2847 & \zkroundtime{43.412} & \zkroundtime{3.692} & \zkroundtime{1525.429} & \zkroundtime{153.954}\\
    \hline
    \texttt{dhcp\_15\_4\_10} & \zkroundspace{8704} & - & OOT & - & - & - & 3693 & \zkroundtime{57.808} & \zkroundtime{4.507} & \zkroundtime{1945.531} & \zkroundtime{196.874}\\
    \texttt{dhcp\_32\_5\_10} & \zkroundspace{21120} & - & - & - & - & - & 4689 & \zkroundtime{73.016} & \zkroundtime{7.661} & \zkroundtime{2472.423} & \zkroundtime{251.296}\\
    \hhline{|=|=|=|=|=|=|=|=|=|=|=|=|}
    \texttt{rr\_2} & \zkroundspace{18} & \zkroundspace{378} & \zkroundtime{0.33} & \zkroundtime{0.03} & \zkroundtime{0.01} & \zkroundtime{0.03} & 678 & \zkroundtime{7.323} & \zkroundtime{0.71} & \zkroundtime{114.109} & \zkroundtime{14.377}\\
    \texttt{rr\_3} & \zkroundspace{81} & \zkroundspace{9873} & \zkroundtime{3.148} & \zkroundtime{0.14} & \zkroundtime{0.07} & \zkroundtime{0.06} & 3791 & \zkroundtime{47.535} & \zkroundtime{4.085} & \zkroundtime{1058.592} & \zkroundtime{116.297}\\
    \texttt{rr\_4} & \zkroundspace{324} & \zkroundspace{159396} & \zkroundtime{83.567} & \zkroundtime{1.6} & \zkroundtime{0.73} & \zkroundtime{0.33} & 13746 & \zkroundtime{202.964} & \zkroundtime{35.628} & \zkroundtime{4090.545} & \zkroundtime{468.305} \\
    \texttt{rr\_5} & \zkroundspace{1215} & \zkroundspace{2007375} & \zkroundtime{2000.242} & \zkroundtime{21.694} & \zkroundtime{11.843} & \zkroundtime{3.342} & 38427 & \zkroundtime{651.650} & \zkroundtime{267.770} & OOM & -\\
     \hline
    \end{tabular}
    \label{tab:experimental_results}
\end{table*}

Folding leaves the range conditions of \cref{eq:fold-families} untouched.
A folded witness is essentially uniform in $\mathbb{Z}_p$ and would fail any range
check, so we state every range proof on the per-obligation commitments, where
$\mathsf{zkrp}$ applies unchanged and the checks on $\delta_i$ aggregate across all
obligations into a single Bulletproof~\cite{bunz2018bulletproofs}, with the
verifier deriving the commitment to $\delta_i$ homomorphically as before.
Separating the two layers also preserves the integer encoding of
\cref{sec:primitives}: folding argues purely over $\mathbb{Z}_p$, while the lift to
$\mathbb{Z}$ stays per obligation.

\begin{theorem}[Soundness of Folding]\label{thm:folding}
Let the commitment scheme be binding except with probability
$\varepsilon_{\mathsf{bind}}$, let $\mathsf{zkmm}$, $\mathsf{zkeq}$ and
$\mathsf{zkrp}$ be knowledge sound with knowledge errors
$\varepsilon_{\mathsf{zkmm}}$, $\varepsilon_{\mathsf{zkeq}}$ and
$\varepsilon_{\mathsf{zkrp}}$, let $C$ satisfy \cref{eq:schedule}, and let
$\hat n' M^2 + \hat n M_G M + M + 1 < p$, where $M_G$ bounds the entries of
$G_\pub$ and $h_\pub$, and $\hat n'$, $\hat n$ are the dimensions of $\lambda_i$, $\mu_i$.
Then, from any accepting prover, one can extract for every $i \in [N]$ integer
witnesses bound to the per-obligation commitments that satisfy
\cref{eq:farkas-zk-proof}, except with probability
${\cal O}(\max_i c_i / p) + \varepsilon_{\mathsf{zkmm}} +
\varepsilon_{\mathsf{zkeq}} + \varepsilon_{\mathsf{zkrp}} + \varepsilon_{\mathsf{bind}}$.
\end{theorem}

The wraparound condition is per obligation and independent of $N$, which enters
only through the Schwartz--Zippel term; we present the proof in
\Cref{sec:foldingsecurity}.

\paragraph{Complexity}
Folding replaces the $N$ invocations of $\mathsf{zkmm}$ and $\mathsf{zkeq}$ by
one each.
It costs $4|K|$ group elements in the certificate, ${\cal O}(N + |K|)$
exponentiations for the verifier, and $N(N-1)$ pairwise products for the prover.
Because $|K|$ grows faster than $N$, the schedule also bounds how far a single
fold scales, and the batch size becomes a tuning knob: folding a task in batches
of $B$ obligations trades $\lceil N/B \rceil$ invocations against
${\cal O}(N B^{0.59})$ cross-term commitments.

\section{Experimental Evaluation}\label{sec:eval}
We implemented an end-to-end prototype\footnote{\artifactlink} of both the explicit-state and symbolic algorithms and use it to investigate three research questions. \textbf{(RQ1)} Can either scheme verify realistic protocols within practical time and memory budgets? \textbf{(RQ2)} How do the two schemes compare in scalability on these protocols (explicit-state in terms of state-space size, symbolic in terms of the number of proof obligations)? \textbf{(RQ3)} In which settings is each scheme preferable?

Our main finding is that the two approaches are complementary. The explicit-state scheme performs well on small systems with negligible setup cost but degrades quadratically in $|S|$. By contrast, the symbolic scheme scales to systems with up to $2^{20.9}$ states and $2^{13.7}$ proof obligations within a two-hour budget, albeit with a higher constant overhead.

\paragraph{Setup}
Our front-end includes a toy certifier input language for guarded-command programs, B\"uchi automata, and piecewise-linear ranking functions, and computes batches and obligations, using  Z3~\cite{DBLP:conf/tacas/MouraB08} to derive the Farkas witnesses. Our front-end is written in Python; the zero-knowledge protocols are written in Rust and receive their input in JSON format.
The explicit-state back-end builds on the Arkworks~\cite{arkworks} implementation of
KZG10: it interpolates the membership polynomials $p_{S_0}$ and $p_T$ by inverse
FFT and proves each of them on its full evaluation set with a single batched opening,
whose quotient is likewise computed via FFT.
The symbolic back-end instantiates $\mathsf{zkmm}$ with zkMatrix~\cite{CongYY24}
and $\mathsf{zkrp}$ with a \texttt{bls12-381} Bulletproofs library~\cite{bls_bulletproofs}.
The prover computes the commitments to the secret matrices
$A_{\sf s}^{\sf T}$ and vectors $-b_{\sf s}^{\sf T}$ (see \cref{eq:dual}), along with
their $\mathsf{zkrp}$ range proofs, once per distinct matrix and reuses them across
obligations; the verifier caches correspondingly. Obligations are discharged
concurrently across cores, in batches of 200 to bound the memory footprint.
A separate folding back-end implements the construction of \Cref{sec:folding} and
folds all obligations of a task in one shot ($B = N$).
We ran our benchmarks on AWS EC2 \texttt{c6g.8xlarge} instances (32 vCPU, 64GiB RAM, ARM Graviton2) running Ubuntu 26.04.

\paragraph{Benchmarks}
We evaluate on three communication and coordination protocols, each parameterised so that the
state space and the number of obligations scale independently, which lets us stress-test the
two back-ends along separate axes.
The benchmark \texttt{exb\_i$\langle d\rangle$a$\langle a\rangle$}, presented in
\Cref{fig:backoff}, has initial delay $d \in \{1,\dots,128\}$ and maximum attempt count
$a \in \{2,3,4\}$; the state space grows with $d \cdot a$ while the number of obligations
depends only on $a$, and we prove that the protocol does not wait indefinitely.
The benchmark \texttt{dhcp\_$\langle$w$_1\rangle$\_$\langle$a$\rangle$\_$\langle$w$_2\rangle$}
models a client requesting a network address~\cite[p.~13]{rfc2131}, with wait bounds
$w_1$ and $w_2$ and at most $a$ attempts in its exponential-backoff loop; variants tagged
\texttt{noOFF} omit a non-essential intermediate state, and we prove that the client either
fails or is configured correctly without waiting indefinitely.
The benchmark \texttt{rr\_$\langle k\rangle$} models a $k$-process round-robin scheduler whose
state space and number of obligations both grow rapidly with $k$, making it the most demanding
family, and we prove that the scheduler is starvation-free.

\paragraph{Metrics}
For each instance we report the problem size $|S|$, with the embedding size
$\sum_\star |E_\star|$ for the explicit-state scheme and the obligation count for the
symbolic scheme, and four timings covering the pipeline end-to-end: \emph{enumeration} produces the JSON input
(set enumeration and embedding for the explicit-state scheme; Farkas witness extraction via
Z3 for the symbolic scheme), \emph{setup} generates the KZG or SRS parameters, \emph{prover}
performs commitment and proof generation, and \emph{verifier} checks the proof.

\paragraph{Discussion}
\Cref{tab:experimental_results} shows the results.
Below $2^{11.8}$ states, the explicit-state scheme is preferable: it has lower setup, prover, and verifier times (e.g., \texttt{exb\_i4a3}: $2$\,s vs.\ $1$\,min for the symbolic prover), and its
polynomial commitment yields a constant-size certificate.
This regime fits early-stage protocol design and hardware controllers.
For larger systems, the symbolic scheme is the only viable option: it proves and verifies a
$2^{20.9}$-state DHCP instance (\texttt{dhcp\_10000\_3\_10}) in under half an hour, while the
explicit-state scheme times out in enumeration alone.
The two schemes also differ in what they assume: the explicit-state algorithms make no
assumption on the structure of the model, whereas the symbolic algorithm requires systems
to be expressed as linear guarded commands.
The two schemes are therefore complementary, trading expressivity for performance.

\section{Related Work}
Prior work at the intersection of zero-knowledge proofs and automated reasoning has proposed protocols for theorem proving with different logical backends: zkUNSAT for propositional unsatisfiability~\cite{LuoAHPTW22}, zkSMT for first-order unsatisfiability in decidable theories~\cite{LuickKAHPPT0L24}, and zkPi for interactive theorem proving~\cite{DBLP:conf/ccs/LauferOB24,DBLP:conf/cade/Moura021}.
None of these approaches performs model checking: they establish the validity of public logical statements with a secret proof witness, whereas we verify a secret system against a temporal specification, so that the theorems themselves (the obligations of \cref{thm:ranking}) contain secret components. Moreover, our approach mandates commitments to the system as a core component, preventing a malicious prover from replacing the system under analysis, whether during, before, or after the proof.
Privacy-preserving verification has also been explored through multi-party computation, with solutions for CTL model checking and runtime verification~\cite{DBLP:conf/wpes/JudsonLAP20,DBLP:conf/ccs/HenzingerKT25}; to the best of our knowledge, ours is the first implementation using zero-knowledge proofs.

The proof rule of \Cref{thm:ranking} builds on results on fair termination in nondeterministic B\"uchi automata~\cite{Vardi91,DBLP:journals/iandc/GrumbergFR85}. Our schemes extend to alternative proof rules: generalised B\"uchi automata~\cite{DBLP:journals/tcs/KupfermanV05}, other temporal specifications~\cite{DBLP:conf/popl/CookGPRV07,DBLP:conf/tacas/CookKP15}, and, in principle, probabilistic model checking~\cite{DBLP:conf/tacas/DimitrovaFHM16,DBLP:conf/cav/AbateGR25,DBLP:conf/cav/AbateGR24,DBLP:conf/cav/HenzingerMSZ25,DBLP:journals/corr/abs-2512-00270,lics26}
and cyber-physical systems~\cite{DBLP:conf/hybrid/PrajnaJ04,DBLP:conf/cdc/PrajnaJP04}; the continuous domains of the latter two, however, entail a generalisation of our cryptographic routines and remain future work.

Linear programming duality, of which Farkas' lemma is a special case, is a standard technique in ranking-function synthesis and verification~\cite{DBLP:conf/cav/ColonSS03,DBLP:conf/vmcai/PodelskiR04}, and zero-knowledge proofs for linear-algebraic relations are well studied~\cite{bunz2018bulletproofs,CongYY24}; our approach is the first to use duality as the structural backbone of a zero-knowledge proof.

Folding admits alternative realisations: a recursive binary-tree fold in
the style of Nova~\cite{nova} scales better with the number of obligations at the
cost of multiple folding rounds, whereas a multi-instance Lagrange-basis fold as
in ProtoGalaxy~\cite{protogalaxy} also uses a single challenge but incurs higher
prover overhead; we adopt the progression-free one-shot fold of
\cref{sec:folding} for its simplicity.
The underlying relation (\cref{eq:farkas-zk-proof}) is also SNARK-friendly and
could be instantiated directly with Groth16 or
PlonK~\cite{DBLP:conf/eurocrypt/Groth16,DBLP:journals/iacr/GabizonWC19},
achieving constant-size proofs and efficient verification at the cost of
increased prover overhead; our tailored sub-protocols resolve this trade-off in
favour of the prover. In particular, the transformation behind $\mathsf{zkrp}$,
which lifts batched Bulletproofs range proofs on Pedersen
commitments~\cite{bunz2018bulletproofs} to the two-tier matrix commitment,
integrates readily with existing Bulletproofs implementations and may be of
independent interest.

\section{Conclusion}
We have presented the first approach to model checking in zero knowledge, enabling formal verification in scenarios where the confidentiality of the system is crucial.
Our implementation of the explicit-state and symbolic schemes, with its one-shot folding of proof obligations, has demonstrated the practical viability of our new technology. Our work raises many scientific questions, from the security implications related to commitment before and after a model checking proof, to the information a public specification may reveal about a secret system.

\section*{Acknowledgments}
This work was supported in part by the Amazon Research Award {\em Neural Software Verification} (Fall 2024), and by the Advanced Research and Invention Agency (ARIA) under the {\em Safeguarded AI} programme.

\bibliographystyle{plainurl}
\bibliography{biblio}

@inproceedings{FiatS86,
  author       = {Amos Fiat and
                  Adi Shamir},
  editor       = {Andrew M. Odlyzko},
  title        = {How to Prove Yourself: Practical Solutions to Identification and Signature
                  Problems},
  booktitle    = {Advances in Cryptology - {CRYPTO} '86, Santa Barbara, California,
                  USA, 1986, Proceedings},
  series       = {Lecture Notes in Computer Science},
  pages        = {186--194},
  publisher    = {Springer},
  year         = {1986}
}

@inproceedings{DBLP:conf/ccs/LauferOB24,
  author       = {Evan Laufer and
                  Alex Ozdemir and
                  Dan Boneh},
  title        = {{zkPi}: Proving {Lean} Theorems in Zero-Knowledge},
  booktitle    = {{CCS}},
  pages        = {4301--4315},
  publisher    = {{ACM}},
  year         = {2024}
}

@inproceedings{ChaumP92,
  author       = {David Chaum and
                  Torben Pryds Pedersen},
  title        = {Wallet Databases with Observers},
  booktitle    = {CRYPTO},
  series       = {Lecture Notes in Computer Science},
  volume       = {740},
  pages        = {89--105},
  publisher    = {Springer},
  year         = {1992}
}

@inproceedings{LuoAHPTW22,
  author       = {Ning Luo and
                  Timos Antonopoulos and
                  William R. Harris and
                  Ruzica Piskac and
                  Eran Tromer and
                  Xiao Wang},
  title        = {Proving {UNSAT} in Zero Knowledge},
  booktitle    = {{CCS}},
  pages        = {2203--2217},
  publisher    = {{ACM}},
  year         = {2022}
}

@inproceedings{LuickKAHPPT0L24,
  author       = {Daniel Luick and
                  John C. Kolesar and
                  Timos Antonopoulos and
                  William R. Harris and
                  James Parker and
                  Ruzica Piskac and
                  Eran Tromer and
                  Xiao Wang and
                  Ning Luo},
  title        = {{ZKSMT:} {A} {VM} for Proving {SMT} Theorems in Zero Knowledge},
  booktitle    = {{USENIX} Security Symposium},
  publisher    = {{USENIX} Association},
  year         = {2024}
}

@article{Vardi91,
  author       = {Moshe Y. Vardi},
  title        = {Verification of Concurrent Programs: The Automata-Theoretic Framework},
  journal      = {Ann. Pure Appl. Log.},
  volume       = {51},
  number       = {1-2},
  pages        = {79--98},
  year         = {1991}
}

@article{Lamport83,
  author       = {Leslie Lamport},
  title        = {Specifying Concurrent Program Modules},
  journal      = {{ACM} Trans. Program. Lang. Syst.},
  volume       = {5},
  number       = {2},
  pages        = {190--222},
  year         = {1983}
}

@inproceedings{AlurH96,
  author       = {Rajeev Alur and
                  Thomas A. Henzinger},
  title        = {Reactive Modules},
  booktitle    = {{LICS}},
  pages        = {207--218},
  publisher    = {{IEEE} Computer Society},
  year         = {1996}
}

@book{KatzLindell2014,
  author       = {Jonathan Katz and
                  Yehuda Lindell},
  title        = {Introduction to Modern Cryptography, Second Edition},
  publisher    = {{CRC} Press},
  year         = {2014}
}

@inproceedings{KateZG10,
  author       = {Aniket Kate and
                  Gregory M. Zaverucha and
                  Ian Goldberg},
  title        = {Constant-Size Commitments to Polynomials and Their Applications},
  booktitle    = {{ASIACRYPT}},
  series       = {Lecture Notes in Computer Science},
  volume       = {6477},
  pages        = {177--194},
  publisher    = {Springer},
  year         = {2010}
}

@misc{rfc793,
    series =    {Request for Comments},
    number =    793,
    howpublished =  {RFC 793},
    publisher = {RFC Editor},
    doi =       {10.17487/RFC0793},
    url =       {https://www.rfc-editor.org/info/rfc793},
    author =    {Postel, Jon},
    title =     {{Transmission Control Protocol}},
    pagetotal = 91,
    year =      1981,
    month =     sep,
}

@misc{rfc2131,
    series =    {Request for Comments},
    number =    2131,
    howpublished =  {RFC 2131},
    publisher = {RFC Editor},
    doi =       {10.17487/RFC2131},
    url =       {https://www.rfc-editor.org/info/rfc2131},
    author =    {Ralph Droms},
    title =     {{Dynamic Host Configuration Protocol}},
    pagetotal = 45,
    year =      1997,
    month =     mar,
}

@software{arkworks,
  author = {arkworks contributors},
  title = {\texttt{arkworks} {zkSNARK} ecosystem},
  url = {https://arkworks.rs},
  year = {2022},
}

@inproceedings{CongYY24,
  author       = {Mingshu Cong and
                  Tsz Hon Yuen and
                  Siu{-}Ming Yiu},
  title        = {zkMatrix: Batched Short Proof for Committed Matrix Multiplication},
  booktitle    = {AsiaCCS},
  publisher    = {{ACM}},
  year         = {2024}
}

@software{bls_bulletproofs,
    author = {numerous contributors},
    title = {\texttt{bls\_bulletproofs}},
    url = {https://github.com/maidsafe/bls_bulletproofs},
    year = {2022}
}

@article{farkas1902theorie,
  title={Theorie der einfachen Ungleichungen.},
  author={Farkas, Julius},
  journal={Journal f{\"u}r die reine und angewandte Mathematik (Crelles Journal)},
  volume={1902},
  number={124},
  pages={1--27},
  year={1902},
  publisher={De Gruyter Berlin, New York}
}

@inproceedings{bunz2018bulletproofs,
  title={Bulletproofs: Short proofs for confidential transactions and more},
  author={B{\"u}nz, Benedikt and Bootle, Jonathan and Boneh, Dan and Poelstra, Andrew and Wuille, Pieter and Maxwell, Greg},
  booktitle={2018 IEEE symposium on security and privacy (SP)},
  pages={315--334},
  year={2018},
  organization={IEEE}
}

@inproceedings{couteau2021efficient,
  author       = {Geoffroy Couteau and
                  Michael Kloo{\ss} and
                  Huang Lin and
                  Michael Reichle},
  title        = {Efficient Range Proofs with Transparent Setup from Bounded Integer
                  Commitments},
  booktitle    = {{EUROCRYPT} {(3)}},
  series       = {Lecture Notes in Computer Science},
  volume       = {12698},
  pages        = {247--277},
  publisher    = {Springer},
  year         = {2021}
}

@inproceedings{hader2024smt,
  title={An SMT-LIB Theory of Finite Fields},
  author={Hader, Thomas and Ozdemir, Alex},
  booktitle={Proceedings of the 22nd International Workshop on Satisfiability Modulo Theories (SMT 2024), Montreal, Canada, July, 22-23, 2024.},
  volume={3725},
  year={2024},
  organization={CEUR Workshop Proceedings}
}

@inproceedings{DBLP:conf/focs/Pnueli77,
  author       = {Amir Pnueli},
  title        = {The Temporal Logic of Programs},
  booktitle    = {{FOCS}},
  pages        = {46--57},
  publisher    = {{IEEE} Computer Society},
  year         = {1977}
}

@book{DBLP:books/daglib/0020348,
  author       = {Christel Baier and
                  Joost{-}Pieter Katoen},
  title        = {Principles of model checking},
  publisher    = {{MIT} Press},
  year         = {2008}
}

@book{DBLP:books/daglib/0007403-2,
  author       = {Edmund M. Clarke and
                  Orna Grumberg and
                  Daniel Kroening and
                  Doron A. Peled and
                  Helmut Veith},
  title        = {Model checking, 2nd Edition},
  publisher    = {{MIT} Press},
  year         = {2018}
}

@book{alur2015principles,
  title={Principles of cyber-physical systems},
  author={Alur, Rajeev},
  year={2015},
  publisher={MIT press}
}

@inproceedings{DBLP:conf/podc/MannaP89,
  author       = {Zohar Manna and
                  Amir Pnueli},
  title        = {A Hierarchy of Temporal Properties},
  booktitle    = {{PODC}},
  pages        = {377--410},
  publisher    = {{ACM}},
  year         = {1990}
}

@inproceedings{DBLP:conf/lics/VardiW86,
  author       = {Moshe Y. Vardi and
                  Pierre Wolper},
  title        = {An Automata-Theoretic Approach to Automatic Program Verification (Preliminary
                  Report)},
  booktitle    = {{LICS}},
  pages        = {332--344},
  publisher    = {{IEEE} Computer Society},
  year         = {1986}
}

@inproceedings{DBLP:conf/popl/CookGPRV07,
  author       = {Byron Cook and
                  Alexey Gotsman and
                  Andreas Podelski and
                  Andrey Rybalchenko and
                  Moshe Y. Vardi},
  title        = {Proving that programs eventually do something good},
  booktitle    = {{POPL}},
  pages        = {265--276},
  publisher    = {{ACM}},
  year         = {2007}
}

@inproceedings{DBLP:conf/tacas/CookKP15,
  author       = {Byron Cook and
                  Heidy Khlaaf and
                  Nir Piterman},
  title        = {Fairness for Infinite-State Systems},
  booktitle    = {{TACAS}},
  series       = {Lecture Notes in Computer Science},
  pages        = {384--398},
  publisher    = {Springer},
  year         = {2015}
}

@article{DBLP:journals/iandc/GrumbergFR85,
  author       = {Orna Grumberg and
                  Nissim Francez and
                  Johann A. Makowsky and
                  Willem P. de Roever},
  title        = {A Proof Rule for Fair Termination of Guarded Commands},
  journal      = {Inf. Control.},
  volume       = {66},
  number       = {1/2},
  pages        = {83--102},
  year         = {1985}
}

@inproceedings{DBLP:conf/nips/GiacobbeKPT24,
  author       = {Mirco Giacobbe and
                  Daniel Kroening and
                  Abhinandan Pal and
                  Michael Tautschnig},
  title        = {Neural Model Checking},
  booktitle    = {NeurIPS},
  year         = {2024}
}

@inproceedings{DBLP:conf/lics/BurchCMDH90,
  author       = {Jerry R. Burch and
                  Edmund M. Clarke and
                  Kenneth L. McMillan and
                  David L. Dill and
                  L. J. Hwang},
  title        = {Symbolic Model Checking: $10^{20}$ States and Beyond},
  booktitle    = {{LICS}},
  pages        = {428--439},
  publisher    = {{IEEE} Computer Society},
  year         = {1990}
}

@incollection{DBLP:reference/mc/Holzmann18,
  author       = {Gerard J. Holzmann},
  title        = {Explicit-State Model Checking},
  booktitle    = {Handbook of Model Checking},
  pages        = {153--171},
  publisher    = {Springer},
  year         = {2018}
}

@inproceedings{DBLP:conf/tacas/ColonS01,
  author       = {Michael Col{\'{o}}n and
                  Henny Sipma},
  title        = {Synthesis of Linear Ranking Functions},
  booktitle    = {{TACAS}},
  series       = {Lecture Notes in Computer Science},
  volume       = {2031},
  pages        = {67--81},
  publisher    = {Springer},
  year         = {2001}
}

@inproceedings{DBLP:conf/tacas/UrbanGK16,
  author       = {Caterina Urban and
                  Arie Gurfinkel and
                  Temesghen Kahsai},
  title        = {Synthesizing Ranking Functions from Bits and Pieces},
  booktitle    = {{TACAS}},
  series       = {Lecture Notes in Computer Science},
  volume       = {9636},
  pages        = {54--70},
  publisher    = {Springer},
  year         = {2016}
}

@inproceedings{DBLP:conf/vmcai/PodelskiR04,
  author       = {Andreas Podelski and
                  Andrey Rybalchenko},
  title        = {A Complete Method for the Synthesis of Linear Ranking Functions},
  booktitle    = {{VMCAI}},
  series       = {Lecture Notes in Computer Science},
  volume       = {2937},
  pages        = {239--251},
  publisher    = {Springer},
  year         = {2004}
}

@book{DBLP:series/txtcs/KroeningS16,
  author       = {Daniel Kroening and
                  Ofer Strichman},
  title        = {Decision Procedures - An Algorithmic Point of View, Second Edition},
  series       = {Texts in Theoretical Computer Science. An {EATCS} Series},
  publisher    = {Springer},
  year         = {2016}
}

@incollection{DBLP:series/faia/BarrettSST09,
  author       = {Clark W. Barrett and
                  Roberto Sebastiani and
                  Sanjit A. Seshia and
                  Cesare Tinelli},
  title        = {Satisfiability Modulo Theories},
  booktitle    = {Handbook of Satisfiability},
  series       = {Frontiers in Artificial Intelligence and Applications},
  volume       = {185},
  pages        = {825--885},
  publisher    = {{IOS} Press},
  year         = {2009}
}

@article{DBLP:journals/cacm/MouraB11,
  author       = {Leonardo Mendon{\c{c}}a de Moura and
                  Nikolaj S. Bj{\o}rner},
  title        = {Satisfiability modulo theories: introduction and applications},
  journal      = {Commun. {ACM}},
  volume       = {54},
  number       = {9},
  pages        = {69--77},
  year         = {2011}
}

@inproceedings{DBLP:conf/cade/Moura021,
  author       = {Leonardo de Moura and
                  Sebastian Ullrich},
  title        = {The {Lean} 4 Theorem Prover and Programming Language},
  booktitle    = {{CADE}},
  series       = {Lecture Notes in Computer Science},
  volume       = {12699},
  pages        = {625--635},
  publisher    = {Springer},
  year         = {2021}
}

@inproceedings{DBLP:conf/cav/Duret-LutzRCRAS22,
  author       = {Alexandre Duret{-}Lutz and
                  Etienne Renault and
                  Maximilien Colange and
                  Florian Renkin and
                  Alexandre Gbaguidi Aisse and
                  Philipp Schlehuber{-}Caissier and
                  Thomas Medioni and
                  Antoine Martin and
                  J{\'{e}}r{\^{o}}me Dubois and
                  Cl{\'{e}}ment Gillard and
                  Henrich Lauko},
  title        = {From Spot 2.0 to Spot 2.10: What's New?},
  booktitle    = {{CAV} {(2)}},
  series       = {Lecture Notes in Computer Science},
  volume       = {13372},
  pages        = {174--187},
  publisher    = {Springer},
  year         = {2022}
}

@inproceedings{DBLP:conf/tacas/MouraB08,
  author       = {Leonardo Mendon{\c{c}}a de Moura and
                  Nikolaj S. Bj{\o}rner},
  title        = {{Z3:} An Efficient {SMT} Solver},
  booktitle    = {{TACAS}},
  series       = {Lecture Notes in Computer Science},
  volume       = {4963},
  pages        = {337--340},
  publisher    = {Springer},
  year         = {2008}
}

@inproceedings{DBLP:conf/asiacrypt/BonehLS01,
  author       = {Dan Boneh and
                  Ben Lynn and
                  Hovav Shacham},
  editor       = {Colin Boyd},
  title        = {Short Signatures from the Weil Pairing},
  booktitle    = {Advances in Cryptology - {ASIACRYPT} 2001, 7th International Conference
                  on the Theory and Application of Cryptology and Information Security,
                  Gold Coast, Australia, December 9-13, 2001, Proceedings},
  series       = {Lecture Notes in Computer Science},
  pages        = {514--532},
  publisher    = {Springer},
  year         = {2001},
  url          = {https://doi.org/10.1007/3-540-45682-1\_30},
  doi          = {10.1007/3-540-45682-1\_30},
  bibsource    = {dblp computer science bibliography, https://dblp.org}
}

@article{DBLP:journals/dam/GalbraithPS08,
  author       = {Steven D. Galbraith and
                  Kenneth G. Paterson and
                  Nigel P. Smart},
  title        = {Pairings for cryptographers},
  journal      = {Discret. Appl. Math.},
  volume       = {156},
  number       = {16},
  pages        = {3113--3121},
  year         = {2008},
  url          = {https://doi.org/10.1016/j.dam.2007.12.010},
  doi          = {10.1016/J.DAM.2007.12.010},
  bibsource    = {dblp computer science bibliography, https://dblp.org}
}

@inproceedings{DBLP:conf/crypto/Pedersen91,
  author       = {Torben P. Pedersen},
  editor       = {Joan Feigenbaum},
  title        = {Non-Interactive and Information-Theoretic Secure Verifiable Secret
                  Sharing},
  booktitle    = {Advances in Cryptology - {CRYPTO} '91, 11th Annual International Cryptology
                  Conference, Santa Barbara, California, USA, August 11-15, 1991, Proceedings},
  series       = {Lecture Notes in Computer Science},
  pages        = {129--140},
  publisher    = {Springer},
  year         = {1991},
  url          = {https://doi.org/10.1007/3-540-46766-1\_9},
  doi          = {10.1007/3-540-46766-1\_9},
  bibsource    = {dblp computer science bibliography, https://dblp.org}
}

@inproceedings{DBLP:conf/crypto/BellareG92,
  author       = {Mihir Bellare and
                  Oded Goldreich},
  editor       = {Ernest F. Brickell},
  title        = {On Defining Proofs of Knowledge},
  booktitle    = {Advances in Cryptology - {CRYPTO} '92, 12th Annual International Cryptology
                  Conference, Santa Barbara, California, USA, August 16-20, 1992, Proceedings},
  series       = {Lecture Notes in Computer Science},
  pages        = {390--420},
  publisher    = {Springer},
  year         = {1992},
  url          = {https://doi.org/10.1007/3-540-48071-4\_28},
  doi          = {10.1007/3-540-48071-4\_28},
  bibsource    = {dblp computer science bibliography, https://dblp.org}
}

@inproceedings{DBLP:conf/cav/ColonSS03,
  author       = {Michael Col{\'{o}}n and
                  Sriram Sankaranarayanan and
                  Henny Sipma},
  editor       = {Warren A. Hunt Jr. and
                  Fabio Somenzi},
  title        = {Linear Invariant Generation Using Non-linear Constraint Solving},
  booktitle    = {Computer Aided Verification, 15th International Conference, {CAV}
                  2003, Boulder, CO, USA, July 8-12, 2003, Proceedings},
  series       = {Lecture Notes in Computer Science},
  pages        = {420--432},
  publisher    = {Springer},
  year         = {2003},
  url          = {https://doi.org/10.1007/978-3-540-45069-6\_39},
  doi          = {10.1007/978-3-540-45069-6\_39},
  bibsource    = {dblp computer science bibliography, https://dblp.org}
}

@inproceedings{DBLP:conf/eurocrypt/Groth16,
  author       = {Jens Groth},
  editor       = {Marc Fischlin and
                  Jean{-}S{\'{e}}bastien Coron},
  title        = {On the Size of Pairing-Based Non-interactive Arguments},
  booktitle    = {Advances in Cryptology - {EUROCRYPT} 2016 - 35th Annual International
                  Conference on the Theory and Applications of Cryptographic Techniques,
                  Vienna, Austria, May 8-12, 2016, Proceedings, Part {II}},
  series       = {Lecture Notes in Computer Science},
  pages        = {305--326},
  publisher    = {Springer},
  year         = {2016},
  url          = {https://doi.org/10.1007/978-3-662-49896-5\_11},
  doi          = {10.1007/978-3-662-49896-5\_11},
  bibsource    = {dblp computer science bibliography, https://dblp.org}
}

@article{DBLP:journals/iacr/GabizonWC19,
  author       = {Ariel Gabizon and
                  Zachary J. Williamson and
                  Oana Ciobotaru},
  title        = {{PLONK:} Permutations over Lagrange-bases for Oecumenical Noninteractive
                  arguments of Knowledge},
  journal      = {{IACR} Cryptol. ePrint Arch.},
  volume       = {2019},
  pages        = {953},
  year         = {2019},
  url          = {https://eprint.iacr.org/2019/953},
  bibsource    = {dblp computer science bibliography, https://dblp.org}
}

@inproceedings{DBLP:conf/wpes/JudsonLAP20,
  author       = {Samuel Judson and
                  Ning Luo and
                  Timos Antonopoulos and
                  Ruzica Piskac},
  title        = {Privacy Preserving {CTL} Model Checking through Oblivious Graph Algorithms},
  booktitle    = {WPES@CCS},
  pages        = {101--115},
  publisher    = {{ACM}},
  year         = {2020}
}

@inproceedings{DBLP:conf/ccs/HenzingerKT25,
  author       = {Thomas A. Henzinger and
                  Mahyar Karimi and
                  K. S. Thejaswini},
  title        = {Privacy-Preserving Runtime Verification},
  booktitle    = {{CCS}},
  pages        = {2774--2787},
  publisher    = {{ACM}},
  year         = {2025}
}

@article{DBLP:journals/tcs/KupfermanV05,
  author       = {Orna Kupferman and
                  Moshe Y. Vardi},
  title        = {From complementation to certification},
  journal      = {Theor. Comput. Sci.},
  volume       = {345},
  number       = {1},
  pages        = {83--100},
  year         = {2005}
}

@inproceedings{DBLP:conf/tacas/DimitrovaFHM16,
  author       = {Rayna Dimitrova and
                  Luis Mar{\'{\i}}a Ferrer Fioriti and
                  Holger Hermanns and
                  Rupak Majumdar},
  title        = {Probabilistic CTL\({}^{\mbox{*}}\): The Deductive Way},
  booktitle    = {{TACAS}},
  series       = {Lecture Notes in Computer Science},
  pages        = {280--296},
  publisher    = {Springer},
  year         = {2016}
}

@inproceedings{DBLP:conf/cav/AbateGR25,
  author       = {Alessandro Abate and
                  Mirco Giacobbe and
                  Diptarko Roy},
  title        = {Quantitative Supermartingale Certificates},
  booktitle    = {{CAV} {(2)}},
  series       = {Lecture Notes in Computer Science},
  pages        = {3--28},
  publisher    = {Springer},
  year         = {2025}
}

@inproceedings{DBLP:conf/hybrid/PrajnaJ04,
  author       = {Stephen Prajna and
                  Ali Jadbabaie},
  title        = {Safety Verification of Hybrid Systems Using Barrier Certificates},
  booktitle    = {{HSCC}},
  series       = {Lecture Notes in Computer Science},
  pages        = {477--492},
  publisher    = {Springer},
  year         = {2004}
}

@inproceedings{DBLP:conf/cdc/PrajnaJP04,
  author       = {Stephen Prajna and
                  Ali Jadbabaie and
                  George J. Pappas},
  title        = {Stochastic safety verification using barrier certificates},
  booktitle    = {{CDC}},
  pages        = {929--934},
  publisher    = {{IEEE}},
  year         = {2004}
}

@inproceedings{lics26,
  author       = {Alessandro Abate and
                  Mirco Giacobbe and
                  Sergey Ichtchenko and
                  Diptarko Roy},
  title        = {Complete {\(\omega\)}-Regular Supermartingale Certificates},
  booktitle    = {{LICS}},
  series       = {LIPIcs},
  volume       = {380},
  pages        = {3:1--3:28},
  publisher    = {Schloss Dagstuhl - Leibniz-Zentrum f{\"{u}}r Informatik},
  year         = {2026}
}

@inproceedings{DBLP:conf/cav/AbateGR24,
  author       = {Alessandro Abate and
                  Mirco Giacobbe and
                  Diptarko Roy},
  title        = {Stochastic Omega-Regular Verification and Control with Supermartingales},
  booktitle    = {{CAV} {(3)}},
  series       = {Lecture Notes in Computer Science},
  pages        = {395--419},
  publisher    = {Springer},
  year         = {2024}
}

@inproceedings{DBLP:conf/cav/HenzingerMSZ25,
  author       = {Thomas A. Henzinger and
                  Kaushik Mallik and
                  Pouya Sadeghi and
                  Dorde Zikelic},
  title        = {Supermartingale Certificates for Quantitative Omega-Regular Verification
                  and Control},
  booktitle    = {{CAV} {(2)}},
  series       = {Lecture Notes in Computer Science},
  pages        = {29--55},
  publisher    = {Springer},
  year         = {2025}
}

@article{DBLP:journals/corr/abs-2512-00270,
  author       = {Satoshi Kura and
                  Hiroshi Unno},
  title        = {A Hierarchy of Supermartingales for {\(\omega\)}-Regular Verification},
  journal      = {CoRR},
  volume       = {abs/2512.00270},
  year         = {2025}
}

@inproceedings{nova,
  author       = {Abhiram Kothapalli and
                  Srinath Setty and
                  Ioanna Tzialla},
  title        = {Nova: Recursive Zero-Knowledge Arguments from Folding Schemes},
  booktitle    = {{CRYPTO}},
  pages        = {359--388},
  year         = {2022}
}

@misc{protogalaxy,
  author       = {Liam Eagen and
                  Ariel Gabizon},
  title        = {{ProtoGalaxy}: Efficient {ProtoStar}-style Folding of Multiple Instances},
  howpublished = {Cryptology ePrint Archive, Paper 2023/1106},
  year         = {2023},
  url          = {https://eprint.iacr.org/2023/1106}
}

@article{schwartz1980,
  author       = {Jacob T. Schwartz},
  title        = {Fast Probabilistic Algorithms for Verification of Polynomial Identities},
  journal      = {Journal of the {ACM}},
  volume       = {27},
  number       = {4},
  pages        = {701--717},
  year         = {1980}
}

@inproceedings{zippel1979,
  author       = {Richard Zippel},
  title        = {Probabilistic Algorithms for Sparse Polynomials},
  booktitle    = {{EUROSAM}},
  pages        = {216--226},
  year         = {1979}
}

@article{erdos1936,
  author       = {Paul Erd{\H{o}}s and
                  Paul Tur{\'a}n},
  title        = {On Some Sequences of Integers},
  journal      = {Journal of the London Mathematical Society},
  volume       = {11},
  number       = {4},
  pages        = {261--264},
  year         = {1936}
}

@article{salem1942,
  author       = {Rapha{\"e}l Salem and
                  Donald C. Spencer},
  title        = {On Sets of Integers Which Contain No Three Terms in Arithmetical
                  Progression},
  journal      = {Proceedings of the National Academy of Sciences},
  volume       = {28},
  number       = {12},
  pages        = {561--563},
  year         = {1942}
}

@article{behrend1946,
  author       = {Felix A. Behrend},
  title        = {On Sets of Integers Which Contain No Three Terms in Arithmetical
                  Progression},
  journal      = {Proceedings of the National Academy of Sciences},
  volume       = {32},
  number       = {12},
  pages        = {331--332},
  year         = {1946}
}

\appendix

\section{Open Science}

We provide four tools and prototypes, as well as a series of benchmark data files, that enable readers to evaluate and reproduce the paper's core contributions. We provide:
\begin{itemize}
    \item {\tt encoder}: A lightweight preprocessing front-end tool written in Python which implements a toy certifier input language for guarded-command programs, B\"uchi automata, and piecewise-linear ranking functions, and computes batches and obligations, using  Z3~\cite{DBLP:conf/tacas/MouraB08} to derive the Farkas witnesses. This tool has built-in benchmarking capabilities to time explicit and symbolic output generation.
    \item {\tt zkmc-explicit}: A prototype implementation of our explicit approach, written in Rust, which takes output from {\tt encoder} as input and provides library methods to prove and verify as per our specification. This prototype comes with a benchmark tool that times setup, proving, and verifying of a given input.
    \item {\tt zkmc-symbolic/non-folding}: A prototype implementation of our symbolic approach, written in Rust, which takes output from {\tt encoder} as input and provides library methods to prove and verify as per our specification. The prototype comes with a benchmark tool that times setup, proving, and verifying of a given input.
    \item {\tt zkmc-symbolic/folding}: A prototype implementation of our symbolic folding approach, written in Rust, which takes output from {\tt encoder} as input and provides library methods to prove and verify as per our specification. The prototype comes with a benchmark tool that times setup, proving, and verifying of a given input.
    \item Example guarded command files: Various models that correspond to our benchmarked instances in guarded command form, which can be readily used with {\tt encoder} to build files which can then be used as input into our prototypes. These files consist of: a minimal {\tt example.gc}; a simple {\tt counter} from 0 to {\tt maxVal}; our {\tt round-robin.gc} example used in benchmarking; three of our models of exponential backoff, namely {\tt exp\_backoff\_state\_opt\_small.gc}, \\{\tt exp\_backoff\_state\_opt.gc}, and \\{\tt exp\_backoff\_guard\_opt.gc}; and our model of DHCP, named {\tt dhcp.gc}.
\end{itemize}
We provide a singular repository consisting of all code and benchmark files, split by tool (with benchmarks and examples inside {\tt encoder}), with full instructions on how to run any artifacts within their own folder structure. This repository can be found at \artifactlink.

\section{Proof Sketch of \texorpdfstring{$\mathsf{zkrp}$}{zkrp}}\label{sec:zkrpsecurity}
Completeness holds by construction.
We argue knowledge soundness in the algebraic group model (AGM) augmented with a random oracle (ROM), and zero knowledge in the random oracle model.

\paragraph{Soundness.}
Given an accepting proof $\pi$, we construct an extractor $\mathcal{E}$ that recovers a witness for the relation $\mathcal{R}_{\mathsf{zkrp}}$.
The extractor first invokes the Bulletproofs extractor on $(\{c_i\}, \pi_{\mathsf{Bullet}})$ to obtain per-entry openings $(v_i, r_i)$ satisfying $c_i = g^{v_i} h^{r_i}$ and $0 \le v_i \le M$.
The extractor then recovers the matrix-level blinding factor $r_A$.
Under the AGM, the prover outputs $\hat c_A$ together with its algebraic representation over the public pairing bases $\{e(g, g')^{\alpha^k}\}$ and $e(g, h')$, and the extractor reads $r_A$ directly as the coefficient of $e(g, h')$ in this representation.
What remains is to show that $(\{v_i\}, \{r_i\}, r_A)$ is consistent with $\hat c_A$.
Substituting the algebraic representation of $\hat c_A$ into the verification equation reduces it to the polynomial identity $f(\alpha) - f(z) = (\alpha - z) \cdot q(\alpha)$ at the challenge $z = H(\hat c_A \| c_1 \| \cdots \| c_l)$, where $f$ is the matrix polynomial defined by $\{v_i\}$.
Since the random oracle distributes $z$ uniformly, the Schwartz--Zippel lemma implies coefficient-wise equality of the two polynomials with overwhelming probability.
Therefore $\hat c_A$ commits exactly to the matrix defined by $\{v_i\}$ with blinding $r_A$, and the extractor outputs the witness $(\{v_i\}, \{r_i\}, r_A)$.

\paragraph{Zero Knowledge.}
We construct a simulator $\mathcal{S}$ that, given only $\hat c_A$, produces transcripts indistinguishable from those of an honest prover.
We grant $\mathcal{S}$ the setup trapdoors $\alpha$ and $\beta$, which is a standard simulation assumption.
On input $\hat c_A$, the simulator proceeds as follows:
\begin{itemize}
  \item Sample $c_i \in \mathbb{G}_1$ uniformly for $i = 1, \dots, l$,  and invoke the Bulletproofs simulator to obtain   $\pi_{\mathsf{Bullet}}$.
  \item Compute the challenge $z = H(\hat c_A \| c_1 \| \cdots \| c_l)$ and the aggregate $c = \prod_{i=1}^{l} c_i^{z^{i-1}}$.
  \item Sample $\pi_{\mathsf{eq}} \in \mathbb{G}_1$ uniformly, and define $\theta = \bigl(\hat c_A \cdot e(c, g')^{-1} \bigm/  e(\pi_{\mathsf{eq}}, (g')^{\alpha - z})\bigr)^{1/\beta}$, which makes the verification equation hold by construction.
  \item Output $\pi = (\{c_i\}, \pi_{\mathsf{eq}}, \theta, \pi_{\mathsf{Bullet}})$.
\end{itemize}
Each component of the simulated transcript is indistinguishable from its real counterpart.
The Pedersen commitments $c_i$ are perfectly hiding, so the simulated $c_i$ have the same distribution as the honest ones.
The Bulletproofs simulator produces range proofs computationally indistinguishable from real ones.
Finally, $(\pi_{\mathsf{eq}}, \theta)$ is uniform among the pairs that satisfy the verification equation, matching the honest prover, who randomises $\pi_{\mathsf{eq}}$ via the blinding scalar $\mu$ sampled in $\mathsf{zkrp.Prove}$ (\cref{fig:ZKMC-S2-routines}) and absorbs the resulting offset into $\theta$.
Hence no probabilistic polynomial-time distinguisher separates the simulated transcript from a real one, and $\mathsf{zkrp}$ is zero knowledge in the random oracle model.

\section{Proof Sketch of Folding}\label{sec:foldingsecurity}

Completeness holds by construction.
The argument for \Cref{thm:folding} separates into a field layer and an integer
layer, which do not interact.

\paragraph{Field Layer.}
The binding property of the commitment scheme and the knowledge soundness of
$\mathsf{zkmm}$ and $\mathsf{zkeq}$ yield openings of the folded instances, which
the homomorphic fold maps back to openings of the per-obligation commitments.
\Cref{sec:folding} established that under \cref{eq:schedule} no diagonal residual
shares a monomial with a freely chosen cross-term aggregate.
A prover that violates some obligation therefore faces a non-zero polynomial of
degree at most $2 \max_i c_i$ in the challenge, which the Schwartz--Zippel lemma
bounds as above; we argue each of the four families separately and add the bounds.
Up to this term, the extracted openings satisfy \cref{eq:fold-families} over
$\mathbb{Z}_p$ for every $i$ individually.
No step here assigns an integer meaning to a value, so the wraparound of the folded
witnesses is immaterial.

\paragraph{Integer Layer.}
The range proofs are stated on the per-obligation commitments, so by binding they
constrain the very same openings, entry-wise:
$\lambda_i, \mu_i, \delta_i \in [0, M]$ and, after removing the offsets,
$A_i, b_i \in [-M, M]$.
Fix an obligation $i$ and evaluate both sides of
$A_i \lambda_i = \alpha_i = G_i \mu_i$ over $\mathbb{Z}$ under the canonical
signed lift of \cref{sec:primitives}.
The two sides are congruent modulo $p$ and bounded in magnitude by
$\hat n' M^2$ and $\hat n M_G M$ respectively, so their difference is smaller than
$p$ and they are equal as integers.
Applying the same argument to
$\delta_i = b_i \lambda_i + h_i \mu_i - 1 \in [0, M]$ yields
$b_{\sec,i}^\T \lambda_{\sec,i} + h_{\pub,i}^\T \mu_{\sec,i} \leq -1$ over
$\mathbb{Z}$.
Together these are the conditions of \cref{eq:farkas-zk-proof}.
Both bounds are per obligation, which is why the condition of
\Cref{thm:folding} does not depend on $N$.

\paragraph{Zero Knowledge.}
Zero knowledge is inherited from the sub-protocols.
The cross-term commitments are perfectly hiding and are the only additional
message the prover sends, and the folded instances are opened solely through
$\mathsf{zkmm}$, $\mathsf{zkeq}$ and $\mathsf{zkrp}$, each run with independent
randomness.

\end{document}